\documentclass[prb,aps,amsmath,amssymb,amsfonts,floatfix,superscriptaddress,
showpacs,twocolumn]{revtex4}
\usepackage{graphicx}
\usepackage[dvips]{color}
\usepackage{dcolumn}
\def\bavs3{BaVS$_3$}
\def\srvo3{SrVO$_3$}
\def\t2g{$t_{2g}$}
\def\eg{$e_g$}

\def\A1g{$A_{1g}$}

\begin{document}

\title{
Dynamical Mean-Field Theory within an Augmented Plane-Wave Framework:\\
Assessing Electronic Correlations in the Iron Pnictide LaFeAsO
}
\author{Markus~Aichhorn}
\affiliation{Centre de Physique Th\'eorique, \'Ecole Polytechnique, CNRS,
91128 Palaiseau Cedex, France}
\author{Leonid~Pourovskii}
\affiliation{Centre de Physique Th\'eorique, \'Ecole Polytechnique, CNRS,
91128 Palaiseau Cedex, France}
\author{Veronica~Vildosola}
\affiliation{Centre de Physique Th\'eorique, \'Ecole Polytechnique, CNRS,
91128 Palaiseau Cedex, France}
\affiliation{Departamento de F\'{\i}sica, Comisi\'{o}n Nacional de Energ\'{\i}a At\'{o}mica (CNEA-CONICET),
Provincia de Buenos Aires, San Mart\'{\i}n, 1650, Argentina}
\affiliation{Japan Science and Technology Agency, CREST, Kawaguchi 332-0012, Japan}
\author{Michel~Ferrero}
\affiliation{Institut de Physique Th{\'e}orique, CEA/DSM/IPhT-CNRS/URA 2306 CEA-Saclay,
F-91191 Gif-sur-Yvette, France}
\affiliation{Centre de Physique Th\'eorique, \'Ecole Polytechnique, CNRS,
91128 Palaiseau Cedex, France}
\author{Olivier~Parcollet}
\affiliation{Institut de Physique Th{\'e}orique, CEA/DSM/IPhT-CNRS/URA 2306 CEA-Saclay,
F-91191 Gif-sur-Yvette, France}
\author{Takashi~Miyake}
\affiliation{Research Institute for Computational Sciences, AIST, Tsukuba 305-8568, Japan}
\affiliation{Japan Science and Technology Agency, CREST, Kawaguchi
  332-0012, Japan}
\affiliation{Japan Science and Technology Agency, TRIP, Kawaguchi 
332-0012, Japan}
\author{Antoine~Georges}
\affiliation{Centre de Physique Th\'eorique, \'Ecole Polytechnique, CNRS,
91128 Palaiseau Cedex, France}
\affiliation{Coll\`ege de France, 11 place Marcelin Berthelot,
75231 Paris Cedex 05, France}
\affiliation{Japan Science and Technology Agency, CREST, Kawaguchi 332-0012, Japan}
\author{Silke Biermann}
\affiliation{Centre de Physique Th\'eorique, \'Ecole Polytechnique, CNRS,
91128 Palaiseau Cedex, France}
\affiliation{Japan Science and Technology Agency, CREST, Kawaguchi 332-0012, Japan}

\begin{abstract}
We present an approach that combines the local density approximation (LDA) and the
dynamical mean-field theory (DMFT) in the framework of the full-potential linear
augmented plane waves (FLAPW) method. Wannier-like functions for the
correlated shell are constructed by  projecting local orbitals onto a set of
Bloch eigenstates located within a certain energy window.
The screened Coulomb interaction and Hund's coupling are calculated
from a first-principle constrained RPA scheme.
We apply this LDA+DMFT implementation, in conjunction with a continuous-time
quantum Monte-Carlo algorithm, to the study of electronic correlations in LaFeAsO.
Our findings support the physical picture of a metal with intermediate correlations.
The average value of the mass renormalization of the Fe 3$d$ bands is about $1.6$,
in reasonable agreement with the picture inferred from photoemission experiments.
The discrepancies between different LDA+DMFT calculations (all technically
correct) which have been reported in the literature are shown to have two causes:
i) the specific value of the interaction parameters used in these calculations and
ii) the degree of localization of the Wannier orbitals chosen to represent the
Fe 3$d$ states, to which many-body terms are applied.
The latter is a fundamental issue in the application of many-body calculations,
such as DMFT, in a realistic setting.
We provide strong evidence that the DMFT approximation is more accurate and more
straightforward to implement when well-localized orbitals are constructed
from a large energy window encompassing Fe-3$d$, As-4$p$ and O-2$p$, and point
out several difficulties associated with the use of extended Wannier functions
associated with the low-energy iron bands. Some of these issues have important
physical consequences, regarding in particular the sensitivity to the
Hund's coupling.
\end{abstract}

\maketitle

\section{Introduction}

This article has two purposes.
The first one is to present a new implementation of dynamical mean-field theory
(DMFT) within electronic structure calculation methods.
This implementation is based on a highly precise
full-potential linear augmented plane wave method (FLAPW),
as implemented in the Wien2k electronic structure code~\cite{Wien2k}.
The second purpose of this article is to report on DMFT calculations
for the iron oxypnictide LaFeAsO, the parent compound of the
`$1111$'-family of recently discovered iron-based superconductors.
The strength of electronic correlations in these materials is an
important issue, which has been a subject of debate in
the literature~\cite{haule1,haule2,anisimov2,shorikov1,anisimov3}.

The combination of dynamical mean-field theory with
density-functional theory in the local density
approximation (LDA+DMFT) provides a powerful framework
for the quantitative description of electronic correlations
in a realistic setting.
A number of materials have been
investigated in this framework over the past decade, such as transition metals
and transition-metal oxides, rare-earth and actinide compounds, and
organic conductors.
These examples testify to the progress in
our understanding of the key physical phenomena associated with
the competition between the localized and itinerant characters
of electrons belonging to different orbitals
(see e.g. Refs.~[\onlinecite{geo96,georges_strong,hel02,bie06,
kotliar_dmft_physicstoday,kotliar1}] for reviews).

In the past few years, a new generation of LDA+DMFT implementations
have been put forward~\cite{pav04,lechermann_wannier-bis,ani05,gavri05,sol06,ani06-bis,
amadon_pw_08},
which emphasize the use of Wannier functions
as a natural bridge between the band-structure and the real-space
description of the solid in terms of orbitals.
These functions span the subset of orbitals
which are treated within the many-body DMFT framework.
In this article, we present an implementation of LDA+DMFT within the
full potential linear augmented plane wave (FLAPW) framework,
using atomic orbitals that are promoted to Wannier
functions by a truncated expansion over Bloch functions followed by
an orthonormalization procedure.
This is a simpler alternative to the previous implementation
of DMFT within FLAPW~\cite{lechermann_wannier-bis}, which constructed the
Wannier functions following the prescription of maximal
localisation~\cite{marzari_wannier_1997_prb,sou01}.
The choice of FLAPW is motivated by the high level of accuracy
of this all-electron, full-potential method.
In the present work, we use the Wien2k electronic structure
package~\cite{Wien2k}, and we have constructed an interface to
it that allows for the construction of Wannier-like functions
used in DMFT. Our implementation is described in detail in
Sec.~\ref{sect:theory}.
As a benchmark, we perform calculations on a
test material, SrVO$_3$, which are presented in
Appendix \ref{sect:srvo3} and compared to previously published results
for this material~\cite{lie03,sek04,pav04,pav05,nek05,sol06,nek06}.
Throughout this article, many-body effects are treated in the DMFT framework using
the recently developed continuous-time strong-coupling Quantum Monte
Carlo algorithm of P.Werner and coworkers~\cite{werner_ctqmc,haule_ctqmc_prb_07}.
Because very low temperatures can be reached,
and very high accuracy can be obtained at low-frequency, this algorithm
represents a major computational advance in the field.

In Sec.~\ref{sect:results}, we address the issue of electronic
correlations in LaFeAsO.
The DMFT calculations which have been published soon after the
experimental discovery of superconductivity in the iron oxypnictides
have provided seemingly contradictory answers to this question.
In Refs.~[\onlinecite{haule1,haule2}], K.~Haule and G.~Kotliar proposed
that LaFeAsO is a strongly correlated metal, rather close to the Mott metal-insulator
transition, and characterized by a reduced value of the quasiparticle
coherence scale, resulting in bad metallic behavior.
In contrast, in Refs.~[\onlinecite{anisimov2,shorikov1,anisimov3}],
V.Anisimov and coworkers proposed that these materials are in
a weak to intermediate regime of correlations.

Our LDA+DMFT calculations for LaFeAsO
support the physical picture of a metal with intermediate correlations.
The average value of the mass renormalization of the Fe 3$d$ bands is about $1.6$,
in reasonable agreement with the picture inferred from photoemission experiments.
We also find that there is no technical inconsistency
between different DMFT results reported for LaFeAsO before.
We show that the discrepancies in the literature are due to two causes:
i) the specific value of the interaction parameters used in these calculations and
ii) the degree of localization of the Wannier orbitals chosen to represent the
Fe 3$d$ states, to which many-body terms are applied.

In Sec.~\ref{sect:results}, we perform detailed comparisons between LDA+DMFT
calculations performed with different degree of localization of the correlated
orbitals, associated with different choices of energy
windows for the Wannier construction (and accordingly, different
degrees of screening of the interaction parameters). We
point out several difficulties associated with the
use of more extended Wannier functions associated with the low-energy iron bands only.
Some of these issues have important physical consequences, in particular
regarding the sensitivity to the Hund's coupling.

This article ends with several appendices, reporting on
more detailed aspects or technical issues.
Appendix A is devoted to a benchmark of our implementation
on a ``classical'' test compound, SrVO$_3$.
Appendix B details some technical issues associated with the
projection scheme used to display partial spectral functions
with a given orbital character.
Appendix C discusses the influence of spin-flip
and pair-hopping terms on the degree of correlations, on the basis of
model calculations. We conclude that while these terms are indeed
important close to the Mott transition, they can safely be neglected
in the regime of correlations relevant to LaFeAsO.

\section{Theoretical framework}
\label{sect:theory}

\subsection{Implementation of LDA+DMFT in the APW framework}

\subsubsection{LDA+DMFT in the basis of Bloch waves}

To make this article self-contained and in order to define the main notations,
this subsection begins by briefly reviewing some essential aspects of the
LDA+DMFT framework. The presentation is close to that of
Refs.~[\onlinecite{lechermann_wannier-bis,amadon_pw_08}], where
additional details can be found.

Dynamical mean-field theory is a quantitative method for handling
electron correlations, which can be described as an ``effective atom''
approach.
The self-energy in the solid is approximated by that of a local
model, a generalized Anderson impurity model describing a specific set of
atomic-like orbitals coupled to a self-consistent environment.
The self-consistency requirement is that the local on-site Green's
function of the solid, calculated using this
local self-energy, must coincide with the Green's function of
the effective impurity model.

In order to formulate the local effective atom problem, a set of
(orthonormal) local orbitals $|\chi_m^{\alpha,\sigma}\rangle$,
and corresponding Wannier-like functions
$|w_{\mathbf{k}m}^{\alpha,\sigma}\rangle$, must be constructed.
These Wannier functions span the "correlated" subspace $\mathcal C$ of
the full Hilbert space, in which many-body correlations (beyond LDA)
are taken into account.
This set of orbitals spanning the correlated subspace must be clearly
distinguished from the full basis set
of the problem, in which the
Green's function of the solid can be expressed. Obviously, the basis
set spans a much larger Hilbert space, involving all relevant
electronic shells.

Below, we discuss in details how the
$|w_{\mathbf{k}m}^{\alpha,\sigma}\rangle$ are constructed from
the local orbitals $|\chi_m^{\alpha,\sigma}\rangle$.
The index $m$ is an orbital
index within the correlated subset, $\alpha$ denotes the atom in the
unit cell, and $\sigma$ is the spin degree of freedom. Projections of
quantities of interest on the subset $\mathcal C$ are done using the
projection operator
\begin{equation}
  \hat P^{\alpha,\sigma}(\mathbf k) = \sum_{m\in \mathcal
    C}|w_{\mathbf{k}m}^{\alpha,\sigma}\rangle\langle
  w_{\mathbf{k}m}^{\alpha,\sigma}|.\label{eq:projop}
\end{equation}

The effective impurity model is then constructed for the correlated
subset $\mathcal C$. It is defined by the Green's function of the
effective environment,
$\mathcal{G}^{0,\sigma}_{mm'}(i\omega_n)$ and Hubbard-Kanamori interaction
parameters $U_{mm'm''m'''}$. By solving this model in a suitably
chosen manner one obtains the impurity Green's function
$G_{mm'}^{\sigma,\rm{imp}}(i\omega_n)$ as well as the impurity self-energy
\begin{equation}
  \Sigma_{mm'}^{\sigma,\rm{imp}}(i\omega_n)=\left(\mathcal{G}^{\sigma,0}(i\omega_n)\right)^{-1}_{mm'}
  - \left(G^{\sigma,\rm{imp}}(i\omega_n)\right)^{-1}_{mm'}.\label{eq:sigmaimp}
\end{equation}

For the formulation of the self-consistency condition relating the
lattice Green's function of the solid to the impurity model, it is
convenient to choose the Bloch basis
$|\psi_{\mathbf{k}\nu}^\sigma\rangle $ as the complete basis set of
the problem, since it is a natural output of any electronic structure
calculation. The (inverse) Green's function of the solid expressed in this basis
set is given by:
\begin{equation}
  G^{\sigma}(\mathbf{k},i\omega_{n})^{-1}_{\nu\nu'} =
(i\omega_{n}+\mu-\epsilon_{\mathbf{k}\nu}^\sigma)\delta_{\nu\nu'}-
  \Sigma_{\nu\nu'}^\sigma(\mathbf{k},i\omega_{n}),\label{eq:latt-G}
\end{equation}
where $\epsilon_{\mathbf{k}\nu}^\sigma$ are the Kohn Sham eigenvalues
and $\Sigma_{\nu\nu'}^\sigma(\mathbf{k},i\omega_{n})$ is the approximation to
the self-energy obtained by the solution of the DMFT impurity
problem. It is obtained by "upfolding" the impurity local
self-energy as
\begin{equation}
  \Sigma_{\nu\nu'}^{\sigma}(\mathbf{k},i\omega_{n})
  =\sum_{\alpha,mm'}P_{\nu m}^{\alpha,\sigma*}
  (\mathbf{k})\Delta\Sigma_{mm'}^{\sigma,\rm{imp}}(i\omega_{n})
  P_{m'\nu'}^{\alpha,\sigma}(\mathbf{k}),\label{eq:latt-Self}
\end{equation}
where $P_{m\nu}^{\alpha,\sigma}(\mathbf{k})=\langle
w_{\mathbf{k}m}^{\alpha,\sigma}|\psi_{\mathbf{k},\nu}^{\sigma}\rangle$
are the matrix elements of the projection operator,
Eq.~(\ref{eq:projop}) and
\begin{equation}
\Delta\Sigma_{mm'}^{\sigma,\rm{imp}}(i\omega_{n})=\Sigma_{mm'}^{\sigma,\rm{imp}}(i\omega_{n})-\Sigma_{mm'}^{\rm{dc}}.
\label{sigma_mm}
\end{equation}
Here, $\Sigma_{mm'}^{\sigma,\rm{imp}}$ is the impurity self-energy,
Eq.~(\ref{eq:sigmaimp}), expressed in the local orbitals, and
$\Sigma_{mm'}^{\rm{dc}}$ is a double-counting correction, which will
be discussed in Sect.~\ref{sect:impl_compmeth}.

The local Green's function is obtained by projecting the
lattice Green's function to the set of correlated orbitals $m$ of the
correlated atom $\alpha$ and summing over the full Brillouin zone,
\begin{equation}
  G_{mm'}^{\sigma,\rm{loc}}(i\omega_{n})=
  \sum_{\mathbf{k},\nu\nu'}P_{m\nu}^{\alpha,\sigma}(\mathbf{k})G_{\nu\nu'}^{\sigma}(\mathbf{k},i\omega_n)
  P_{\nu'm'}^{\alpha,\sigma*}(\mathbf{k}).\label{eq:local-G}
\end{equation}
Note that the local quantities
$G_{mm'}^{\sigma,\rm{loc}}(i\omega_{n})$ and
$\Delta\Sigma_{mm'}^{\sigma,\rm{imp}}(i\omega_{n})$ carry also an
index $\alpha$, which we suppressed for better readability.

The self-consistency condition of DMFT imposes that the \emph{local} Green's
function, Eq.~(\ref{eq:local-G}) must coincide with the one obtained from the
effective impurity problem,
\begin{equation}
  \mathbf{G}^{\sigma,\rm{loc}}(i\omega_n)= \mathbf{G}^{\sigma,\rm{imp}}(i\omega_n).\label{eq:SC}
\end{equation}
This equation implies that the Green's function of the effective environment, $\mathcal{G}_0$, must be
self-consistently related to the self-energy of the impurity model through:
\begin{equation}\label{eq:selfconsist-sigma}
\mathcal{G}_0^{-1} = \Sigma_{\rm{imp}} + G_{\rm{loc}}^{-1}
\end{equation}
where the dependence of $G_{\rm{loc}}$ on $\Sigma_{\rm{imp}}$ is specified by
Eqs.~(\ref{eq:latt-G},\ref{eq:local-G}). In practice, the DMFT equations
are solved iteratively: starting from an initial $\mathcal{G}_0$, the impurity model
is solved for $\Sigma_{\rm{imp}}$, and a new $\mathcal{G}_0$ is constructed from
(\ref{eq:selfconsist-sigma}). The cycle is repeated until convergence is reached.

In order to
construct the set of Wannier functions, we start from a set of local
atomic-like orbitals $|\chi_m^{\alpha,\sigma}\rangle$ defined in the unit
cell.
These orbitals can be expanded over the full Bloch basis-set as:
\begin{equation}
|\chi_{\mathbf{k}m}^{\alpha,\sigma}\rangle
=\sum_{\nu}\langle\psi_{\mathbf{k}\nu}^{\sigma}|\chi_{m}^{\alpha,\sigma}\rangle
|\psi_{\mathbf{k}\nu}^{\sigma}\rangle
\end{equation}
This expansion is then truncated by choosing
an energy window $\mathcal W$, and restricting the sum to those Bloch
states with Kohn-Sham energies $\epsilon_{\mathbf{k}\nu}$ within
$\mathcal W$. The number
of bands included in $\mathcal W$
will in general depend on $\mathbf k$ and $\sigma$.
We thus define the modified orbitals (which do not form an
orthonormal set because of the truncation):
\begin{equation}
|\tilde\chi_{\mathbf{k}m}^{\alpha,\sigma}\rangle
=\sum_{\nu\in \mathcal{W}}\langle\psi_{\mathbf{k}\nu}^{\sigma}|\chi_{m}^{\alpha,\sigma}\rangle
|\psi_{\mathbf{k}\nu}^{\sigma}\rangle.\label{eq:Correl-orb-1}
\end{equation}
Let us denote the matrix elements of the projection operator for this
subset as
\begin{equation}
  \widetilde P_{m\nu}^{\alpha,\sigma}(\mathbf{k})=\langle
    \tilde\chi_{m}^{\alpha,\sigma}|\psi_{\mathbf{k}\nu}^{\sigma}\rangle,
    \qquad \nu\in\mathcal{W}
    \label{eq:proj-1}
\end{equation}
The matrix $\widetilde P_{m\nu}^{\alpha,\sigma}(\mathbf{k})$ is not unitary,
except when the sum in (\ref{eq:Correl-orb-1}) is carried over all
Bloch bands. It is also important to note that the matrices
$\mathrm{\widetilde{P}}^{\alpha,\sigma}$ are in general
non-square matrices. They reduce to square matrices only
in the case when the number of Kohn-Sham bands contained
in the chosen window equals at every $\mathbf{k}$-point
the number of correlated local orbitals to be constructed.

The orbitals $\left|\tilde\chi_{\mathbf{k}m}^{\alpha,\sigma}
\right\rangle $ can be orthonormalized, giving a set of
Wannier-like functions:
\begin{equation}
  \left|w_{\mathbf{k}m}^{\alpha,\sigma}\right\rangle
  =\sum_{\alpha',m'}S_{m,m'}^{\alpha,\alpha'}
  \left|\tilde \chi_{\mathbf{k}m'}^{\alpha',\sigma}\right\rangle ,
  \label{eq:wannier}
\end{equation}
where $S_{m,m'}^{\alpha,\alpha'}
=\left\{ O(\mathbf{k},\sigma)^{-1/2}\right\} _{m,m'}^{\alpha,\alpha'}$
and $O_{m,m'}^{\alpha,\alpha'}(\mathbf{k},\sigma)=\left\langle
\tilde \chi_{\mathbf{k}m}^{\alpha,\sigma}\right|\left.\tilde\chi_{\mathbf{k}m'}^{\alpha',\sigma}
\right\rangle $ the overlap matrix elements.

The overlap $O_{m,m'}^{\alpha,\alpha'}(\mathbf{k},\sigma)$ finally reads
\begin{equation}
  O_{m,m'}^{\alpha,\alpha'}(\mathbf{k},\sigma)=\sum_{\mathcal{W}}
  \widetilde{P}_{m\nu}^{\alpha,\sigma}(\mathbf{k})
  \widetilde{P}_{\nu m'}^{\alpha',\sigma*}(\mathbf{k}),\label{eq:over-wannier}
\end{equation}
while the orthonormalized projectors are then written as
\begin{equation}
  P_{m\nu}^{\alpha,\sigma}
  (\mathbf{k})=\underset{\alpha'm'}{\sum}\left
    \{ \left[O(\mathbf{k},\sigma)
    \right]^{-1/2}\right\}_{m,m'}^{\alpha,\alpha'}
  \widetilde{P}_{m'\nu}^{\alpha',\sigma}(\mathbf{k}).
\label{eq:wannier-proj}
\end{equation}

\subsubsection{Augmented plane waves}

In this work, the Bloch basis $|\psi_{\mathbf{k}\nu}^\sigma\rangle $
are expanded in augmented plane waves (APW/LAPW), which will be briefly
described in the following. As was first pointed out by Slater
\cite{Slater}, near atomic nuclei the crystalline potential in a solid
is similar to that of a single atom, while in the region between
nuclei (in the interstitial) the potential is rather smooth and
weakly-varying. Hence, one may introduce a set of basis functions,
{\it augmented plane waves}, adapted to this general shape of the
potential. First the crystal space is divided into non-overlapping
muffin-tin (MT) spheres centered at the atomic sites and the
interstitial region in between. In the interstitial region($I$) the
APW $\phi_{\mathbf{G}}^{\mathbf{k}}(\mathbf{r})$ is simply the
corresponding plane wave for given reciprocal lattice vector
$\mathbf{G}$ and crystal momentum $\mathbf{k}$:
\begin{equation}
  \phi_{\mathbf{G}}^{\mathbf{k}}(\mathbf{r})=\frac{1}{\sqrt{V}}e^{i(\mathbf{k}+\mathbf{G})\mathbf{r}} \quad \mathbf{r}\in I,
\end{equation}
where $V$ is the unit cell volume. This plane wave is augmented
inside each of the MT-spheres by a combination of the radial solutions
of the Schr{\"o}dinger equation in such way that the resulting APW is
continuous at the sphere boundary. The APW are then employed to expand
the Kohn-Sham (KS) eigenstates $\psi_{\mathbf{k}\nu}^{\sigma}(\mathbf{r})$ for the
full Kohn-Sham (KS) potential, (without any shape approximation). In
the original formulation of the APW method the radial solutions
expanding a KS eigenstate inside MT-spheres had to be evaluated at the
corresponding eigenenergy leading to an energy-dependent basis set
and, hence, to a non-linear secular problem. In order to avoid this
complication, linearized versions of the APW method have been
proposed. There are two widely-used schemes for the APW
linearization. In the first, the linear APW (LAPW) method
\cite{LAPWSingh}, the plane wave is augmented within MT-spheres by a
combination of the radial solutions, evaluated at chosen linearization
energies $E_{1l}$, and their energy derivatives. The resulting linear
augmented plane wave then reads:
\begin{widetext}
  \begin{equation}
    \phi_{\mathbf{G}}^{\mathbf{k}}(\mathbf{r})=\left\{ \begin{array}{ll}
        \frac{1}{\sqrt{V}}e^{i(\mathbf{k}+\mathbf{G})\mathbf{r}} & \qquad\mathbf{r}\in I\\
        \underset{lm}{\sum}\left[A_{lm}^{\alpha,\mathbf{k}+\mathbf{G}}u_{l}^{\alpha,\sigma}(r,E_{1l}^{\alpha})+
          B_{lm}^{\alpha,\mathbf{k}+\mathbf{G}}\dot{u}_{l}^{\alpha,\sigma}(r,E_{1l}^{\alpha})\right]Y_{m}^{l}(\hat{r})
        & \qquad\mathbf{r}\in
        R_{MT}^{\alpha}\end{array}\right.\label{eq:LAPW}
  \end{equation}
\end{widetext}
where the index $\alpha=1,...,N_{at}$ runs over all $N_{\alpha}$
atomic sites in the unit cell, the coefficients $A_{lm}$ and $B_{lm}$
are determined from the requirement for the linear APW to be
continuous and differentiable at the sphere boundary, $r$ and
$\hat r$ are the radial and angular parts of the position vector,
respectively. The
energy-independent basis set (\ref{eq:LAPW}) leads to a linear secular
problem, however, compared to the energy-dependent APW,  a larger
number of the LAPW in the basis set is generally required to attain
the same accuracy. In order to decrease the requirement for the number
of APW another linearization scheme, APW+$lo$ \cite{APWlo} has  been
proposed. The APW+$lo$ basis set consists of the augmented plane waves
evaluated at a fixed energy $E_{1l}$:
\begin{equation}
  \phi_{\mathbf{G}}^{\mathbf{k}}(\mathbf{r})=\left\{ \begin{array}{ll}
      \frac{1}{\sqrt{V}}e^{i(\mathbf{k}+\mathbf{G})\mathbf{r}} & \qquad\mathbf{r}\in I\\
      \underset{lm}{\sum}A_{lm}^{\alpha,\mathbf{k}+\mathbf{G}}u_{l}^{\alpha,\sigma}(r,E_{1l}^{\alpha})Y_{m}^{l}(\hat{r})
      & \qquad\mathbf{r}\in
      R_{MT}^{\alpha},\end{array}\right.\label{eq:APW}
\end{equation}
where the coefficient $A_{lm}$ are determined from the requirement for
$\phi_{\mathbf{G}}^{\mathbf{k}}(\mathbf{r})$ to be continuous at the
sphere boundary. To increase the variational freedom of the APW+$lo$
basis set the fixed-energy APW (\ref{eq:APW}) are supplemented for the
physically important orbitals (with $l \leq 3$) by the local orbitals
($lo$) that are not matched to any plane wave in the interstitial and
are defined only within the muffin tin spheres ($\mathbf{r}\in
R_{MT}^{\alpha}$)
\begin{equation}
  \phi_{lm,\alpha}^{lo}(\mathbf{r})=\left[A_{lm}^{\alpha,lo}u_{l}^{\alpha,\sigma}(r,E_{1l}^{\alpha})+
    B_{lm}^{\alpha,lo}\dot{u}_{l}^{\alpha,\sigma}(r,E_{1l}^{\alpha})\right]Y_{m}^{l}(\hat{r})
  \label{eq:lo-apw}
\end{equation}
with the coefficients $A_{lm}$ and $B_{lm}$ chosen from the requirement
of zero value and slope for the local orbital at the sphere boundary.

Additional local orbitals (usually abbreviated with the capital
letters as $LO$, $\phi_{lm,\alpha}^{LO}$) can be introduced to account
for semicore states. They have a similar form as Eq.~(\ref{eq:lo-apw}) with the redefined $A_{lm}^{\alpha,LO}$
and a second term with a coefficient $C_{lm}^{\alpha,LO}$ and the radial function evaluated at a corresponding
energy $E_{2l}^{\alpha}$ for the semicore band. The coefficient $B_{lm}$ is set to $0$ in the APW+$lo$ framework.

Generally, in the full potential augmented plane wave method the LAPW,
APW+$lo$ and $LO$ types of orbitals can be employed simultaneously. The
Kohn-Sham eigenstate is expanded in this mixed basis as:
\begin{equation}
  \psi_{k\nu}^{\sigma}(\mathbf{r})=\overset{N_{b}}
  {\underset{i=1}{\sum}}c_{i\nu}\phi_{i}^{\sigma}(\mathbf{r}),\label{eq:KS-eigen}
\end{equation}
where $N_b$ is the number of the orbitals in the basis set. The
LDA+DMFT framework introduced in the present work can also be
used in conjunction with any mixed APW+$lo$/LAPW/$LO$ basis set.

\subsubsection{Local orbitals and Wannier functions in the APW basis}\label{sect:locorbsinAPW}

Having defined the basis set we may now write down the expression for the Bloch
eigenstate expanded in the general APW basis (\ref{eq:KS-eigen}).
For $\mathbf{r}$ in the interstitial region, it reads:
\begin{equation}
  \psi_{\mathbf{k}\nu}^{\sigma}(\mathbf{r})=\frac{1}{\sqrt{V}}\overset{N_{PW}}
  {\underset{\mathbf{G}}{\sum}}c_{\mathbf{G}}^{\nu,\sigma}(\mathbf{k})e^{i(\mathbf{k}
    +\mathbf{G})\mathbf{r}},
\end{equation}
while for the region within the MT-spheres $\mathbf{r}\in R_{MT}^{\alpha}(\alpha=1,...,N_{at})$, we have:
\begin{widetext}
\begin{equation}
  \begin{split}
    \psi_{\mathbf{k}\nu}^{\sigma}(\mathbf{r})&=\overset{N_{PW}}{\underset{\mathbf{G}}{\sum}}
    c_{\mathbf{G}}^{\nu,\sigma}(\mathbf{k})\underset{lm}{\sum}A_{lm}^{\alpha,\mathbf{k}+
      \mathbf{G}}u_{l}^{\alpha,\sigma}(r,E_{1l}^{\alpha})Y_{m}^{l}(\hat{r})
    +\overset{N_{lo}}{\underset{n_{lo}=1}{\sum}}c_{lo}^{\nu,\sigma}
    \left[A_{lm}^{\alpha,
        lo}u_{l}^{\alpha,\sigma}(r,E_{1l}^{\alpha})+B_{lm}^{\alpha,lo}
      \dot{u}_{l}^{\alpha,\sigma}(r,E_{1l}^{\alpha})\right]Y_{m}^{l}(\hat{r})+\\
    &+\overset{N_{LO}}{\underset{n_{LO}=1}{\sum}}c_{LO}^{\nu,\sigma}
    \left[A_{lm}^{\alpha,
        LO}u_{l}^{\alpha,\sigma}(r,E_{1l}^{\alpha})+C_{lm}^{\alpha,LO}u_{l}^{\alpha,\sigma}(r,E_{2l}^{\alpha})\right]Y_{m}^{l}(\hat{r}),
  \end{split}
  \label{eq:KS-eigen-2}
\end{equation}
\end{widetext}
where $N_{PW}$ is the total number of plane waves considered in the
interstitial which in turn is augmented inside each MT-sphere,
$N_{lo}$ is the number of $\phi_{lm,\alpha}^{lo}(\mathbf{r})$ orbitals of
Eq.~(\ref{eq:lo-apw}) and $N_{LO}$ the corresponding number of
auxiliary orbitals for semicore states
$\phi_{lm,\alpha}^{LO}(\mathbf{r})$.

In the framework of the APW method one has several choices for the `initial' correlated
orbitals $|\chi_m^{\alpha,\sigma}\rangle$.
Any suitable combination of the radial solution of the Schr\"{o}dinger equation and its energy derivative for
a given correlated shell $\{\alpha,l\}$ can be employed, for example,
the $lo$-orbital (\ref{eq:lo-apw}). In the present paper we simply chose the
$|\chi_m^{\alpha,\sigma}\rangle$'s as
the solutions of the Schr\"{o}dinger equation within the MT-sphere
$\left|u_{l}^{\alpha,\sigma}(E_{1l})Y_{m}^{l}\right\rangle$ at the
corresponding linearization energy $E_{1l}$.

Inserting $|\chi_m^{\alpha,\sigma}\rangle=\left|u_{l}^{\alpha,\sigma}(E_{1l})Y_{m}^{l}\right\rangle$
and the expansion (\ref{eq:KS-eigen-2}) of the
Bloch eigenstate in terms of APWs into Eqs.~(\ref{eq:Correl-orb-1},\ref{eq:proj-1}),
and making use of the orthonormality of the radial solutions and their energy derivatives:
\begin{align}
  \left\langle
    u_{l}^{\alpha,\sigma}(E_{1l})Y_{m}^{l}\right.\left|u_{l'}^{\alpha,\sigma}(E_{1l})Y_{m'}^{l'}\right\rangle
  &=\delta_{ll'mm'}
  \label{eq:overlap-uu}\\
  \left\langle u_{l}^{\alpha,\sigma}(E_{1l})Y_{m}^{l}\right.
  \left|\dot{u}_{l'}^{\alpha,\sigma}(E_{1l})Y_{m'}^{l'}
  \right\rangle &=0,
  \label{eq:overlap-uudot}
\end{align}
 one obtains the following expression for the projection operator matrix element:
\begin{equation}
  \begin{split}
    \widetilde P_{m\nu}^{\alpha,\sigma}(\mathbf{k})&=\left\langle u_{l}^{\alpha,\sigma}
      (E_{1l})Y_{m}^{l}\right.\left|\psi_{\mathbf{k}\nu}^{\sigma}
    \right\rangle\\
    &=A_{lm}^{\nu,\alpha}(\mathbf{k},\sigma)
    +\overset{N_{LO}}{\underset{n_{LO}=1}{\sum}}
    C_{lm,LO}^{\nu,\alpha}(\mathbf{k},\sigma),
  \end{split}
  \label{eq:proj-2}
\end{equation}
In this expression, the first term in the right hand side of (\ref{eq:proj-2}) is due to
the contribution from the LAPW and/or APW+$lo$ orbitals
\begin{equation}
  \begin{array}{ll}
    A_{lm}^{\nu,\alpha}(\mathbf{k},\sigma)&=\overset{N_{PW}}
    {\underset{\mathbf{G}}{\sum}}c_{\mathbf{G}}^{\nu,\sigma}(\mathbf{k})
    A_{lm}^{\alpha,\mathbf{k}+\mathbf{G}}\\
    &+\overset{N_{lo}}{\underset{n_{lo}=1}{\sum}}c_{lo}^{\nu,\sigma}
    A_{lm}^{\alpha,lo}+\overset{N_{LO}}{\underset{n_{LO}=1}{\sum}}c_{LO}^{\nu,\sigma}
    A_{lm}^{\alpha,LO}
  \end{array}
  \label{eq:coeff-proj-1}
\end{equation}
and the contribution due to the $LO$ (semicore) orbitals that arises due to mutual non-orthogonality of the
radial solutions of the Schr\"{o}dinger equation for different energies
\begin{equation}
  C_{lm,LO}^{\nu,\alpha}(\mathbf{k},\sigma)=
  c_{LO}^{\nu,\sigma}C_{lm}^{\alpha,LO}
  \tilde{O}_{lm,l'm'}^{\alpha,\sigma},
  \label{eq:coeff-proj-2}
\end{equation}
where $\tilde{O}_{lm,l'm'}^{\alpha,\sigma}$ is the corresponding overlap:
\begin{equation}
  \tilde{O}_{lm,l'm'}^{\alpha,\sigma}=\left\langle u_{l}^{\alpha,\sigma}(E_{1l})Y_{m}^{l}\right.\left|u_{l'}^{\alpha,\sigma}(E_{l,LO})
    Y_{m'}^{l'}\right\rangle \neq 0 \label{eq:overlap-u1u2}
\end{equation}
Then we orthonormalize the obtained local orbitals to form a set
of Wannier-like functions, Eq.~(\ref{eq:wannier}).
The corresponding projection operator matrix elements (\ref{eq:proj-2})
are orthonormalized accordingly using Eq.~(\ref{eq:wannier-proj}).

\subsection{Implementation and computational methods}\label{sect:impl_compmeth}

\subsubsection{FLAPW code}

For the electronic structure calculation we use the full potential APW+$lo$/LAPW code as
implemented in the Wien2k package~\cite{Wien2k}.
We have built an interface that constructs the projectors to the correlated orbitals
($ P_{m\nu}^{\alpha,\sigma}(\mathbf{k})$) out of the
eigenstates produced by the Wien2k code, as described in
Sect.~\ref{sect:locorbsinAPW}.
In order to obtain the local Green's function, the summation over momenta, Eq.~(\ref{eq:local-G}),
is done in the irreducible Brillouin zone (BZ) only, supplemented by a 
symmetrization procedure which is standard in electronic structure calculations,
\begin{equation}
  \sum_{\mathbf k}^{BZ} {\mathbf A}({\mathbf k}) =
  \sum_{s=1}^{N_s}\sum_{\mathbf k}^{IBZ} {\mathcal O}_s{\mathbf
    A}({\mathbf k}){\mathcal O}_s^{\dagger},
\end{equation}
where ${\mathbf A}(\mathbf k)$ is any $\mathbf k$-dependent matrix,
$N_s$ the number of symmetry operations and $\mathcal O_s$ the
symmetrization matrices.
Furthermore, we construct the local orbitals in the local coordinate
system of the corresponding atom. This means that the equivalent atoms
in the unit cell for which the DMFT should be applied, e.g. the two Fe
atoms in the oxypnictides, are exactly the same and the impurity
problem has to be solved only once. Afterward, the
Green's function and self-energies are put back to the global
coordinate system of the crystal in which the Bloch Green's function,
Eq.~(\ref{eq:latt-G}) is formulated.

\subsubsection{Continuous-time quantum Monte-Carlo}

For the solution of the impurity problem we use the strong-coupling
version of the continuous-time
quantum Monte Carlo method (CTQMC)~\cite{werner_ctqmc,haule_ctqmc_prb_07}.
It is based on a hybridization expansion and has proved to be a very efficient solver
for quantum impurity models in the weak and strong correlation
regime.
It allows us to address room temperature
($\beta\equiv 1/kT \approx 40$\,eV$^{-1}$) without problems. In our calculations, we used typically
around $5\cdot 10^6$
Monte-Carlo sweeps and 1000 $\mathbf k$-points in the irreducible BZ.
Since the CTQMC solver computes the Green's function on the imaginary-time axis,
an analytic continuation is needed in order to
obtain results on the real-frequency axis. Here, we choose
to perform a continuation of the impurity self-energy using a
stochastic version of the Maximum Entropy method,\cite{beach_ME}
yielding real an imaginary parts of the retarded self-energy
$\rm{Re}\Sigma(\omega+i0^+),\rm{Im}\Sigma(\omega+i0^+)$ which can be inserted into
Eq.~(\ref{eq:latt-G}) in order to obtain the lattice spectral function
and density of states.

\subsubsection{Many-body interactions}

The CTQMC strong-coupling algorithm can deal with the full rotationally invariant
form of the interaction hamiltonian~\cite{haule_ctqmc_prb_07}.
However, most calculations presented in this article will consider
only the Ising terms of the Hund's coupling, yielding the
interaction Hamiltonian:
\begin{equation}
  \begin{split}
    H_{int}&=\frac{1}{2}\sum_{mm',\sigma}U_{mm'}^{\sigma\sigma}n_{m\sigma}n_{m'\sigma} \\
    &+\frac{1}{2}\sum_{mm'}U_{mm'}^{\sigma\bar{\sigma}}\left(n_{m\uparrow}n_{m'\downarrow}
      + n_{m\downarrow}n_{m'\uparrow}\right),
  \end{split}
\end{equation}
with $U_{mm'}^{\sigma\sigma}$ and $U_{mm'}^{\sigma\bar{\sigma}}$
the reduced interaction matrices for equal and opposite spins,
respectively. This enables us to take advantage of a maximal amount of
conserved quantum numbers and, 
hence, perform the CTQMC calculation without any truncation of the
local basis. The effects of spin-flip and `pair-hopping' terms in the Hund's
interaction are discussed in Appendix~\ref{sect:hundsrule}.

In our approach, the interaction matrices are expressed in terms of
the Slater integrals $F^0$, $F^2$, and $F^4$, where for $d$-electrons
these parameters are related to the Coulomb and Hund's coupling via
$U=F^0$, $J=(F^2+F^4)/14$, and
$F^2/F^4=0.625$.\cite{anisimov_lda+u_review_1997_jpcm} Using standard
techniques the four-index $U$-matrix is calculated, and the reduced 
interaction matrices are then given by $U_{mm'}^{\sigma\bar{\sigma}}=U_{mm'mm'}$ and
$U_{mm'}^{\sigma\sigma}=U_{mm'mm'}-U_{mm'm'm}$. With the above 
definitions, the Coulomb parameters $U$ and $J$ are related to the
matrices via 
\begin{align}
\label{eq:Uaverage}
  U&=\frac{1}{N^2}\sum_{mm'}^N
  U_{mm'}^{\sigma\bar{\sigma}} 
\\
\label{eq:Javerage}
  J&=U-\frac{1}{N(N-1)}\sum_{m\ne m'}^N U_{mm'}^{\sigma\sigma}. 
\end{align}

As mentioned above, the LDA+DMFT scheme (as the LDA+U one) involves a
double-counting correction $\Sigma_{mm'}^{\rm{dc}}$
in Eqs.~(\ref{eq:latt-Self},\ref{sigma_mm}). Indeed, on-site Coulomb interactions are
already treated on mean-field level in LDA. Several forms of the
double-counting correction term have been proposed and
investigated.\cite{anisimov_lda+u_review_1997_jpcm,lichtenstein_magnetism_dmft_2001_prl,Ylvisaker_LSDA+U_2009_prb}
In this work we follow Ref.~\onlinecite{anisimov_lda+u_review_1997_jpcm} and use the
following double-counting correction:
\begin{equation}
  \Sigma_{mm'}^{\sigma,\rm{dc}} =
  \delta_{mm'}\left[ U\left(N_c-\frac{1}{2}\right)- J\left(N_c^\sigma
    -\frac{1}{2}\right)\right],\label{eq:dc}
\end{equation}
where $U$ is the average Coulomb interaction, $J$ the Hund's rule
coupling, $N_c^\sigma$ the spin-resolved occupancy of the correlated
orbitals, and $N_c=N_c^\uparrow+N_c^\downarrow$. We compared the
results obtained with Eq.~(\ref{eq:dc}) also with the double-counting
correction given in Ref.~\onlinecite{held_LDA+DMFT_advphys_2007} which
gave very similar results.

\subsection{Choice of energy window, localization of Wannier functions,
and screening}
\label{sect:energyscales}

The Wannier functions defining the correlated subspace of orbitals for
which a DMFT treatment is performed, are constructed by truncating
the expansion of the initial atomic-like local orbitals to a restricted
energy window $\mathcal{W}$, as described above.
The choice of this energy window is an important issue, which deserves
further discussion.
Indeed, it will determine the shape and the
degree of localization of the resulting Wannier-like functions.

Let us consider first the case of a rather small energy window,
containing only those bands that have dominantly an orbital character
which qualifies them as ``correlated'' (e.g.,
the Fe-$3d$ orbitals in LaFeAsO or the V-$3d$-t$_{2g}$ orbitals in SrVO$_3$).
In that case, the dimension of the Kohn-Sham Hamiltonian used in
Eq.~(\ref{eq:latt-G}) coincides with that of the correlated subspace
$\mathcal{C}$ (i.e. with the number of orbitals involved in the
effective impurity model, for a single correlated atom per cell).
The Wannier-like functions are then quite extended in real space,
and resemble strongly the Wannier orbitals constructed
within other schemes, such as the maximally localized Wannier
construction of Ref.~\onlinecite{marzari_wannier_1997_prb}, or the $N$-th order
muffin-tin basis set downfolded to that set of bands.\cite{and00,and00-2,zur05}
In such a situation, hybridization of the correlated
orbitals with states that lie outside the energy window is
neglected at the DMFT level. Some information about the
hybridization, e.g., of the d-states with ligand orbitals is
of course taken into account through the leakage of the
Wannier orbitals on neighboring ligand sites (see, e.g.,
Ref.~\onlinecite{vildosola1}). 

In contrast, if a larger energy window is chosen, it
will in general contain states treated as correlated as well as
states on which no Hubbard interactions are imposed.
In this case, the dimension of the Kohn-Sham Hamiltonian used in the
LDA+DMFT calculation of the local Green's function
(\ref{eq:latt-G},\ref{eq:local-G}) exceeds the number of
correlated orbitals involved in the effective impurity model.
The Wannier functions are more localized in space,
and the information about the hybridization of the correlated orbitals
with other states within this larger window is carried by the
off-diagonal blocks of the Hamiltonian between correlated and
uncorrelated states.

An instructive case occurs when the correlated bands are
well separated from the uncorrelated bands at all $\mathbf{k}$-points,
but the bands overlap in energy. This situation is realized
for instance for the t$_{2g}$ bands in SrVO$_3$ that extend
into the energy region of the e$_g$ bands.
In order to strictly pick the three correlated t$_{2g}$ bands at
each $\mathbf{k}$-point, one would in that case have to introduce a
$\mathbf{k}$-dependent energy window. For a $\mathbf{k}$-independent
energy window, one will in general have more than three bands at some
$\mathbf{k}$-points, corresponding to some $e_g$ contribution in the
chosen window. This can then be expected to result in
slightly more localized orbitals. An example is given
in Appendix \ref{sect:benchmark}.

\subsection{Local Coulomb interactions, screening and constrained
RPA calculations}
\label{sect:CRPA}

The choice of the energy window influences the value of the
interaction parameters $U_{mm'm''m'''}$ in a crucial manner,
which can be traced back to two main reasons. 

First, the interaction
parameters are related to matrix elements of a screened interaction
between the chosen Wannier functions. The more bands are included in
their construction, the more localized they become
and, hence, the matrix elements increase.
Second, screening effects themselves affect the value of $U$. The more
states are excluded from the screening process, the larger $U$
becomes. In what follows we distinguish carefully between these
two effects.  

In the present work, we apply the present LDA+DMFT implementation to
one of the new high-T$_c$ superconductors, LaFeAsO.
For constructing the Wannier functions,
we focus on an energy window that contains the 10 bands around the
Fermi level with dominantly Fe-$3d$ character and also the bands
coming from the $p$-bands of O and As, which are mainly located in the
energy region $[-6, -2]$\,eV, resulting in a "$dpp$-hamiltonian". In
addition, we performed calculations also for a smaller energy window
containing only the Fe-$3d$ bands, yielding a "$d$-hamiltonian", as
well as for a very large window including around 60 Bloch bands.

The values of the Coulomb interactions $U$ and $J$ are calculated
from the constrained random phase approximation 
(cRPA)\cite{ary04, miyake2}, using the recently developed
scheme for entangled band structures\cite{miyake-disentanglement}.
cRPA calculations for LaFeAsO have been performed before in
Refs.~\onlinecite{nakamura1,miyake1}.
In Ref.~\onlinecite{miyake1} the screened Coulomb
parameters are obtained for three different situations:
(i) by constructing Wannier functions from an energy window comprising
the Fe-$d$ bands only, and screening calculated excluding the
Fe-$d$ channels only,
(ii) by considering a larger window which also includes the As and O-$p$ states,
so that the screening processes, for instance, from the As-$p$ states
to the Fe-$d$ ones are also excluded,
(iii) and finally, a hybrid situation (dubbed `$d$-$dpp$' in Ref.~\onlinecite{miyake1}), in which
the Wannier functions are calculated from an energy window including Fe-$d$, O-$p$, and As-$p$, but 
only the Fe-$d$ states are excluded from the screening. In other
words, the screening is calculated as in (i) and the Wannier functions
as in (ii).
As discussed in Ref.~\onlinecite{miyake1}, this third
option should be appropriate in a situation in which a full
$dpp$-hamiltonian is used but Hubbard interactions are only applied to
the $d$ states. 

Here, we follow the same approach as in the ``$d$-$dpp$'' case
of Ref.~\onlinecite{miyake1}, but using the new disentanglement
scheme of Ref.~\onlinecite{miyake-disentanglement}.
First, a partially screened Coulomb interaction $W_r$ is
constructed as follows:
Wannier functions for the Fe-$d$ states are calculated,
and a basis for the complementary subspace (containing
in particular the ligands, but also higher lying f-states)
is constructed.
Based on the interpolating $d$-band structure, the $P_d$ polarization
is computed, and the partially screened Coulomb interaction
$W_r$ is obtained by screening the
bare Coulomb interaction by all RPA screening processes
{\it except $P_d$}.
Finally, according to the cRPA procedure the Hubbard $U$ matrix 
is composed of the matrix elements of $W_r$ in the basis of
$dpp$-Wannier functions.
As argued in Ref.~\onlinecite{miyake1} this procedure
is suitable for calculations that deal with the full $dpp$-Hamiltonian
in the many-body calculations while explicitly retaining 
Coulomb interactions on the $d$-submanifold only.
In particular, we stress that to the extent that our
projection method produces Wannier functions for the
$dpp$-window, the Hubbard $U$ parameters are expressed in
the same basis as the impurity quantities.

It is important to note that we keep the screening channels in cRPA,
i.e. $P_d$, unchanged when the energy window in our calculation
is varied. Thus, the different energy windows affect only the
localization of the Wannier functions, but not the screening process
of the bare Coulomb interaction, and therefore the effective
interactions are increasing with increasing energy window.
Keeping the screening channels fixed is fully consistent with the
fact that correlations are only included for the $d$ electrons, 
but not for the ligand states.  

For our purposes, we calculate the average Coulomb interaction $U$ and
Hund parameter $J$ from the matrices 
calculated by cRPA. With this $U$ and $J$, the interaction matrices
in the spherical symmetric approximation used in our calculation
are obtained as discussed above. As we will discuss in more detail
below, the comparison of the resulting
$U_{mm'}^{\sigma\sigma}$ and $U^{\sigma\bar{\sigma}}_{mm'}$ with the cRPA matrices shows that for
the $dpp$-hamiltonian, the approximation using atomic values for the
ratios of Slater integrals $F^k$ is well justified, whereas for the
$d$-hamiltonian the cRPA matrices show strong orbital
anisotropies. 

\section{Results for the iron oxypnictide LaFeAsO}
\label{sect:results}

\subsection{Construction of the $dpp$-Hamiltonian}

Let us start the discussion of correlation effects in LaFeAsO
with our results for the $dpp$-hamiltonian, for which Wannier
functions are constructed from the energy window
$\mathcal{W}=[-5.5,2.5]$\,eV. These Wannier functions are
quite well localised. The corresponding Kohn-Sham hamiltonian contains
$22$ Bloch bands, corresponding to the $10$ Fe-$3d$ bands, the $6$
As-$p$ and the $6$ O-$p$ bands.

The local many-body interactions corresponding to this choice of
Wannier functions
are obtained from cRPA, as described in the previous section.
They read:
\begin{align}
  U_{mm'}^{\sigma\sigma}|_{\rm{cRPA}}&=\left(\begin{array}{ccccc}
      0.00 &  1.61 & 1.55 &  2.26 &  2.26 \\
      1.61 &  0.00 & 2.50 &  1.82 &  1.82 \\
      1.55 &  2.50 & 0.00 &  1.70 &  1.70 \\
      2.26 &  1.82 & 1.70 &  0.00 &  1.74 \\
      2.26 &  1.82 & 1.70 &  1.74 &  0.00\end{array}\right)\nonumber\\
  U^{\sigma\bar{\sigma}}_{mm'}|_{\rm{cRPA}}&=\left(\begin{array}{ccccc}
      3.77 &  2.35 & 2.21 &  2.71 &  2.71 \\
      2.35 &  3.94 & 2.87 &  2.44 &  2.44 \\
      2.21 &  2.87 & 3.31 &  2.29 &  2.29 \\
      2.71 &  2.44 & 2.29 &  3.48 &  2.29 \\
      2.71 &  2.44 & 2.29 &  2.29 &  3.48\end{array}\right)\nonumber
\end{align}
The ordering of orbitals in those matrices is $d_{z^2}$,
$d_{x^2-y^2}$, $d_{xy}$, $d_{xz}$, $d_{yz}$.

According to the conventions of the formulae Eqs.~(\ref{eq:Uaverage}),
(\ref{eq:Javerage}),
these matrices correspond to the values: $U=2.69$\,eV and $J=0.79$\,eV.
Using these values of $U$ and $J$, we
construct the spherically symmetric interaction matrices,
\begin{align}
  U_{mm'}^{\sigma\sigma}&=\left(\begin{array}{ccccc}
      0.00 & 1.49 & 1.49 & 2.30 &  2.30 \\
      1.49 & 0.00 & 2.57 & 1.76 &  1.76 \\
      1.49 & 2.57 & 0.00 & 1.76 &  1.76 \\
      2.30 & 1.76 & 1.76 & 0.00 &  1.76 \\
      2.30 & 1.76 & 1.76 & 1.76 &  0.00\end{array}\right)\nonumber\\
  U_{mm'}^{\sigma\bar{\sigma}}&=\left(\begin{array}{ccccc}
      3.59 & 2.19 & 2.19 & 2.73 &  2.73 \\
      2.19 & 3.59 & 2.91 & 2.37 &  2.37 \\
      2.19 & 2.91 & 3.59 & 2.37 &  2.37 \\
      2.73 & 2.37 & 2.37 & 3.59 &  2.37 \\
      2.73 & 2.37 & 2.37 & 2.37 &  3.59\end{array}\right)\nonumber
\end{align}
It is obvious, that the approximation of the cRPA matrices by using
spherical symmetrisation is well justified in this case, with the
largest absolute deviation being $\Delta U\approx 0.35$\,eV, corresponding to
a relative error of around 0.09. The reason for this good agreement is
that in the present case the Wannier functions are already very close
to atomic-like orbitals. We also checked that cRPA yields a
significantly larger value of $U$ for iron-oxide (FeO), as expected
physically.

\subsection{LDA+DMFT Results ($dpp$ hamiltonian)}

\begin{figure}[t]
  \centering
  \includegraphics[width=0.9\columnwidth]{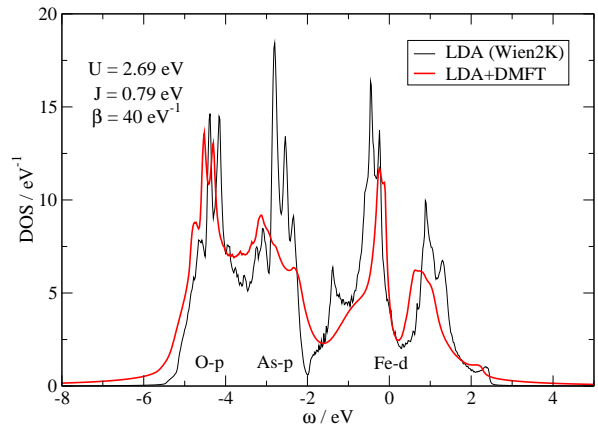}
  \caption{\label{fig:laofeas_dpp_tot}
    (Color online) Total DOS for LaOFeAs, $dpp$ Hamiltonian. Black line: LDA DOS. Red
  line: LDA+DMFT DOS.}
\end{figure}

\begin{figure}[t]
  \centering
  \includegraphics[width=0.9\columnwidth]{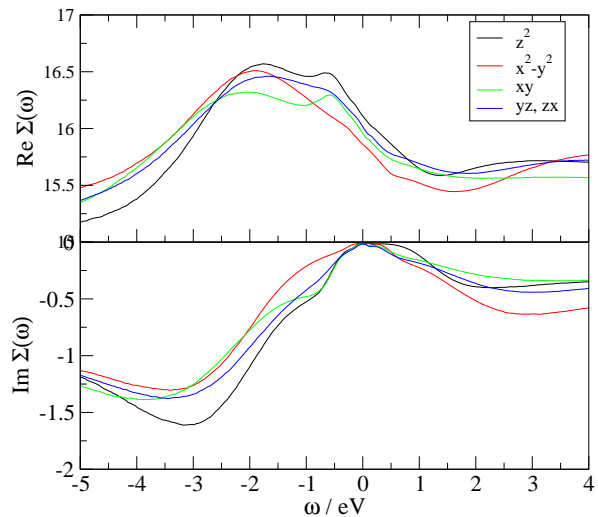}
  \caption{\label{fig:Sigma}
    (Color online) Real (top) and imaginary (bottom) part of the orbital dependent self-energy in the
    $dpp$ Hamiltonian for $U=2.69$, $J=0.79$.}
\end{figure}

\begin{figure}[t]
  \centering
  \includegraphics[width=0.9\columnwidth]{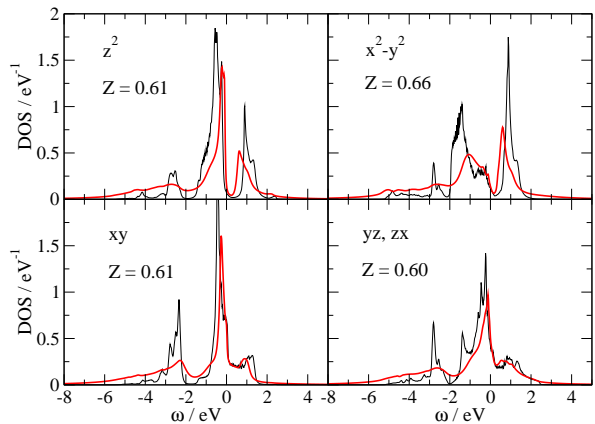}
  \caption{\label{fig:laofeas_dpp_part}
    (Color online) DOS of Fe orbitals in LaOFeAs, $dpp$ Hamiltonian. Color coding as
    in Fig.~\ref{fig:laofeas_dpp_tot}.}
\end{figure}

We carried out LDA+DMFT calculations for the $dpp$-hamiltonian using
the above matrices at an inverse temperature $\beta=40$\,eV$^{-1}$
(room temperature $T=300$K), using the experimental crystal structure
of LaFeAsO.
In Fig.~\ref{fig:laofeas_dpp_tot} we display the resulting total densities of states
(DOS)
together with the corresponding LDA DOS. The total densities of states
were computed from the lattice Green's function, Eq.~(\ref{eq:latt-G}),
traced over all $\nu \in \mathcal{W}$ and integrated over BZ. In order
to obtain the corresponding LDA densities of states
$\Sigma_{\nu\nu'}^{\sigma}(\mathbf{k},i\omega_{n})$ in
Eq.~(\ref{eq:latt-G}) was set to zero.

One sees in Fig. \ref{fig:laofeas_dpp_tot} that
the LDA+DMFT DOS near the Fermi level displays characteristic features of
a metal in an intermediate range of correlations. Both occupied and empty states are
shifted towards the Fermi level due to the Fermi-liquid
renormalizations. 
No high-energy features that would correspond to
lower or upper Hubbard bands are present in the LDA+DMFT electronic
structure.
The Fermi-liquid behavior is clear from the self-energy on the
real-frequency axis, which we plot in Fig.~\ref{fig:Sigma} for the
$dpp$-hamiltonian. Although it shows a quite rich structure as a function of
energy, the real part displays clear linear behavior at low-frequency.
The imaginary part is small around $\omega=0$ and has a quadratic frequency
dependence at low frequency. It does increase to rather large values at higher
frequencies, however, especially for occupied states.
Hence, our results are in general agreement with the
previous calculations of Anisimov {\it et al.}\cite{anisimov2} and
with the experimental photoemission (PES)\cite{malaeb1} and X-ray
absorption (XAS)\cite{kurmaev} spectra of LaFeAsO, which report a
moderately-correlated system with mass renormalisation around $1.8-2.0$.

In order to analyze the strength of correlations for different Fe 3$d$
orbitals we calculated the corresponding quasiparticle residues
$Z_m=\left[1-{\rm Im}[\frac{d\Sigma_{mm}(\omega)}{d\omega}|_{\omega \to
    0}]\right]^{-1}$ from the self-energy Eq.~(\ref{sigma_mm}) on the
Matsubara grid (hence, avoiding all uncertainties related to the
analytical continuation). The values are 0.609, 0.663, 0.609, and
0.596 for the $d_{z^2}$, $d_{x^2-y^2}$, $d_{xy}$, and degenerate
$d_{xz}$/$d_{yz}$ orbitals, resp.
In this $dpp$ energy window the Wannier functions become quite localized
and their spread is expected to be isotropic. Indeed, within
the $dpp$-hamiltonian the difference in $Z_m$ between the orbitals is rather
small.
The resulting value for the average
mass renormalization (between 1.5 and 1.7) is in reasonable agreement
with the experimental estimate of 1.8 extracted in
Ref.~\onlinecite{malaeb1} from experimental PES.
The smaller mass renormalization found in our calculation compared to the
experimental value can be attributed to the single-site approximation
of DMFT. Spatial spin fluctuations, which are completely neglected in
this approach, can eventually increase the effective mass of the
quasiparticles.

The partial densities of states for all Fe
3$d$ orbitals computed within the $dpp$-model are displayed in
Fig.~\ref{fig:laofeas_dpp_part}. The partial LDA+DMFT DOS for the
$x^2-y^2$ and $yz,xz$ orbitals are shifted upwards relative to the
$xy$. Indeed, we found that the crystal field (CF)
splitting between the Fe 3$d$ orbitals is somewhat affected by
correlations. The splitting between the lowest $xy$  and  highest
$z^2$ orbitals remains unchanged ($\approx 0.3$\,eV), while the $x^2-y^2$ and
$yz,xz$ CF levels are shifted upwards by 0.15 and 0.08\,eV relative to
their positions in LDA. In LDA+DMFT they are located at 0.25 and
0.18\,eV, respectively, above the $xy$ orbital.

\begin{figure}[t]
  \centering
  \includegraphics[width=0.9\columnwidth]{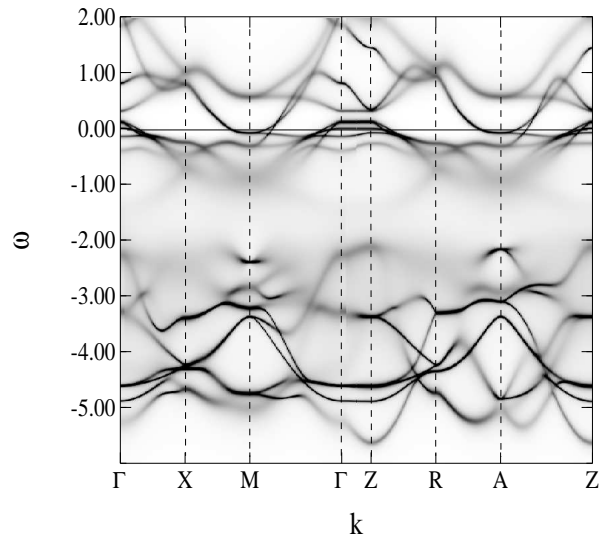}
  \caption{\label{fig:laofeas_dpp_akw}
    Momentum-resolved spectral function of LaOFeAs, $dpp$ Hamiltonian. Dark areas mark large spectral weight.}
\end{figure}

\begin{figure}[t]
  \centering
  \includegraphics[width=0.9\columnwidth]{fig5a.eps}\\
  \includegraphics[width=0.9\columnwidth]{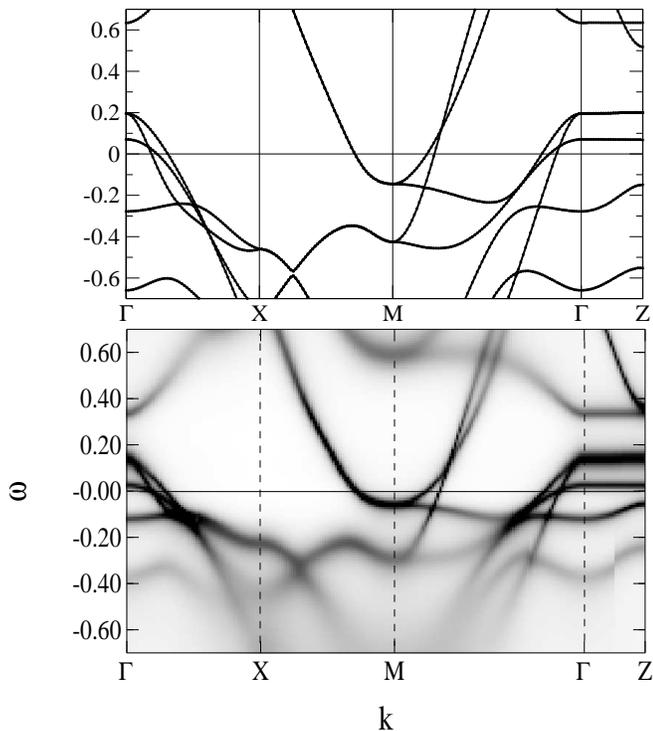}
  \caption{\label{fig:laofeas_dpp_akwzoom}
    Comparison of the momentum-resolved spectral function of LaOFeAs
    at low energies. Upper panel: LDA. Lower panel: LDA+DMFT $dpp$ hamiltonian.} 
\end{figure}

It is also instructive to look at the momentum-resolved spectral
function $A({\mathbf k},\omega)$ of the crystal. It is obtained from
the lattice Green's function, Eq.~(\ref{eq:latt-G}), using the
real-frequency self-energy and tracing over the orbital degrees of
freedom. The result for the $dpp$-model is shown in Fig.~\ref{fig:laofeas_dpp_akw}
for an energy range including Fe-$d$, As-$p$, and O-$p$
states. In agreement with the Fermi-liquid picture of
moderately-correlated quasiparticles discussed above, one can see
well-defined excitations around the Fermi level, which get more
diffuse at higher binding energies. The bands above the Fermi level
are less affected, since the self-energies are quite asymmetric and
smaller for positive frequencies, see
Fig.~\ref{fig:Sigma}. Additionally, it is easy to see that the As-$p$
states, dominantly in the energy range $[-3.5,-2]$\,eV, hybridize
stronger with the Fe-$d$ states and get, thus, affected by
correlations. This effect is almost absent for the O-$p$ states, since
they hybridize much less with Fe-$d$.

In Fig.~\ref{fig:laofeas_dpp_akwzoom} we show a comparison between the LDA band structure and
the LDA+DMFT ${\bf k}$-resolved electronic structure
in a low-energy range around the Fermi
level. This again reveals the coherent quasiparticles at the Fermi
level, as well as more diffuse bands at higher energies.
The crossover between long-lived quasiparticles and more diffuse
states with a shorter lifetime is around $-0.4$\,eV,
in qualitative agreement with existing ARPES data\cite{lu-d-h1}.
A point to mention here is the effect
of the CF splitting on the band structure. For example, a
difference between the LDA and DMFT results can be seen for the
excitation with predominantly $xy$ character. In LDA it forms a
hole-pocket with an excitation energy of $+0.08$\,eV at the $\Gamma$
point. Due to correlations, however, this band is shifted down
significantly to the Fermi level, and the third hole pocket stemming
from the $d_{xy}$ orbital could eventually vanish upon electron
doping.

One has to keep in mind that a direct comparison to experimental data
is difficult for this compound, since (i) the
experiments where done at low temperatures in the SDW phase, whereas
our calculations are done at room-temperature using the tetragonal
crystal structure, and (ii) ARPES experiments on the $1111$ family of pnictide
superconductors are difficult to perform because of difficulties with
single-crystal synthesis. Nevertheless, on a qualitative level,
there is a satisfactory agreement between LDA+DMFT and experiments.

\begin{table}[ht]
  \caption{Quasiparticle weights for different interaction
    parameters, with Wannier orbitals constructed from
    $\mathcal{W}=[-5.5,2.5]$\,eV ($dpp$ hamiltonian). The values in boldface
    correspond to the interaction parameters obtained from cRPA.}
  \begin{tabular}{|c|cccc|}
    \hline
    Interactions & $z^2$ & $x^2-y^2$ & $xy$ & $yz$,$zx$ \\
    \hline
    ${\bf U=2.69}$, ${\bf J=0.79}$    & {\bf 0.61} & {\bf 0.66} & {\bf 0.61} & {\bf 0.60} \\
    $U=2.69$, $J=0.60$    & 0.72 & 0.76 & 0.73 & 0.71 \\
    $U=3.70$, $J=0.80$    & 0.52 & 0.57 & 0.53 & 0.52 \\
    $U=5.00$, $J=0.80$    & 0.41 & 0.45 & 0.43 & 0.42 \\
    \hline
  \end{tabular}
  \label{tabl:Z}
\end{table}

\begin{table}[ht]
  \caption{Quasiparticle weights for different interaction
    parameters, with the Wannier orbitals constructed for a very large
    window $\mathcal{W}=[-5.5,13.6]$\,eV}
  \begin{tabular}{|c|cccc|}
    \hline
    Interactions & $z^2$ & $x^2-y^2$ & $xy$ & $yz$,$zx$ \\
    \hline
    $U=3.00$, $J=0.80$ & 0.62 & 0.66 & 0.58 & 0.58 \\
    $U=3.00$, $J=0.60$ & 0.74 & 0.77 & 0.72 & 0.72 \\
    $U=3.70$, $J=0.80$ & 0.58 & 0.61 & 0.52 & 0.56 \\
    \hline
  \end{tabular}
  \label{tabl:Z2}
\end{table}

We also studied the dependence of the results on the values of the
interaction parameters $U$ and $J$.  The resulting quasi-particle
renormalizations $Z_{m}$ are listed in
Table~\ref{tabl:Z}. Comparing the first two rows, one can see that a
smaller value of $J$ decreases the degree of correlations. The third
line correspond to values similar to the ones used in
Ref.~\onlinecite{anisimov2}, giving very similar results. We also
increased $U$ to the (unphysically) large value of $U=5.0$\,eV, and
the system still displays metallic behavior, although more
correlated.
Hence, our calculations strongly suggest that LaFeAsO
is not close to a Mott metal-insulator transition.

In order to check the robustness of our results, we also investigated
the effect of increasing even further the spatial
localization of the Wannier functions,
corresponding to a very large energy window
$\mathcal{W}=[-5.5,13.6]$\,eV. We did several calculations for
different parameter sets, and the resulting quasi-particle
renormalizations $Z_{m}$ of all these calculations are listed in
Table~\ref{tabl:Z2}. For this case, no cRPA calculations for the
interaction matrices were performed, but it is expected that $U$ and
$J$ will slightly increase with more localised Wannier orbitals. In
that sense, the first row of Table~\ref{tabl:Z2} corresponds to
interaction parameters that could be realised for these Wannier
functions. It
is very satisfying to see the calculations gave almost identical
quasi-particle renormalizations. Also the dependence on $U$ and $J$ is
very similar to the one we found for the $dpp$ hamiltonian. In that
sense we consider our calculations to be converged in terms of the
number of Bloch bands that are included for the construction of the
Wannier functions and the local Hamiltonian.

\subsection{Remarks on calculations using the $d$-Hamiltonian
and extended Wannier functions}

In this section, we address the LDA+DMFT calculations performed with
the so-called $d$-hamiltonian, where only the $10$ Fe-$d$ bands around the
Fermi level are used for the construction of the Wannier orbitals. In doing so,
we shall shed light on the discussion which has appeared in the
literature\cite{haule1,anisimov2,haule2,shorikov1,anisimov3} regarding the
results of LDA+DMFT calculations by different authors, and the
degree of correlations of the $1111$ family of pnictide superconductors.

\subsubsection{Wannier functions and interaction matrices}

\begin{figure}[t]
  \centering
  \includegraphics[width=0.9\columnwidth]{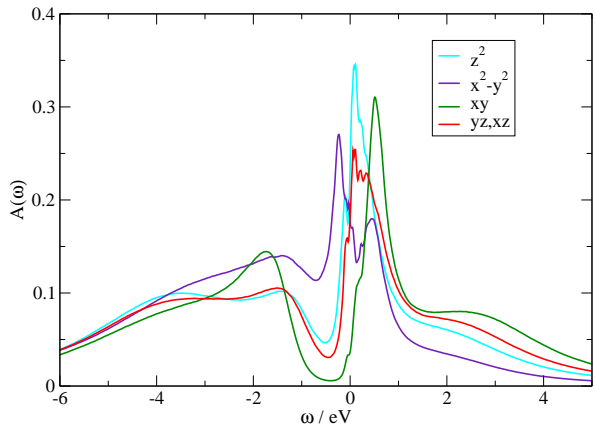}
  \caption{\label{fig:laofeas_d}
    (Color online) Impurity spectral function within the $d$
    hamiltonian using interaction parameters $U=4.0$\,eV and
    $J=0.7$\,eV as have been used in Ref.~\onlinecite{haule1}. For
    those values a very correlated metal is obtained.} 
\end{figure}

The first thing to note is that the Wannier functions constructed from a
small energy window encompassing only the Fe-$d$ bands are quite extended
and very anisotropic, as discussed in details in
Ref.~\onlinecite{vildosola1}. This is directly reflected in the
interaction matrices calculated by cRPA in this restricted energy window:
\begin{align}
  U_{mm'}^{\sigma\sigma}|_{\rm cRPA}&=\left(\begin{array}{ccccc}
      0.00 & 1.41 & 1.26 & 1.87 &  1.87 \\
      1.41 & 0.00 & 1.91 & 1.54 &  1.54 \\
      1.26 &  1.91 &  0.00 &  1.33 &  1.33 \\
      1.87 &  1.54 &  1.33 &  0.00 & 1.44 \\
      1.87 & 1.54 & 1.33 &  1.44 &  0.00
    \end{array}\right)\nonumber\\
  U^{\sigma\bar{\sigma}}_{mm'}|_{\rm cRPA}&=\left(\begin{array}{ccccc}
      3.17 & 2.02 & 1.72 & 2.22 & 2.22 \\
      2.02 & 3.36 & 2.16 & 2.04 &  2.04 \\
      1.72 & 2.16 & 2.17 & 1.73 &  1.73 \\
      2.22 & 2.04 & 1.73 & 2.73 &  1.84 \\
      2.22 & 2.04 & 1.73 & 1.84 & 2.73\end{array}\right)\nonumber
\end{align}
which display a strong orbital dependence. For instance, the intraorbital
(Hubbard) interaction spans from 2.17\,eV to 3.36\,eV. The interaction
matrices in the spherical symmetric approximation using, the averages
$U=2.14$ and $J=0.59$, are
\begin{align}
  U_{mm'}^{\sigma\sigma}&=\left(\begin{array}{ccccc}
      0.00 & 1.25 & 1.25 & 1.85 & 1.85 \\
      1.25 & 0.00 & 2.06 & 1.45 & 1.45 \\
      1.25 & 2.06 & 0.00 & 1.45 & 1.45 \\
      1.85 & 1.45 & 1.45 & 0.00 & 1.45 \\
      1.85 & 1.45 & 1.45 & 1.45 & 0.00
    \end{array}\right)\nonumber\\
  U_{mm'}^{\sigma\bar{\sigma}}&=\left(\begin{array}{ccccc}
      2.82 &  1.77 & 1.77 & 2.18 & 2.18 \\
      1.77 &  2.82 & 2.31 & 1.91 & 1.91 \\
      1.77 &  2.31 & 2.82 & 1.91 & 1.91 \\
      2.18 &  1.91 & 1.91 & 2.82 & 1.91 \\
      2.18 &  1.91 & 1.91 & 1.91 & 2.82\end{array}\right)\nonumber
\end{align}
The largest deviation in this case is $\Delta
U=0.65$\,eV, corresponding to a relative error of about 0.26. This
shows clearly that the spherical approximation is highly questionable
when using only the $d$-bands for the Wannier construction.
Of course, the full anisotropic interaction
matrices can in principle be used in the LDA+DMFT calculation,
but this raises the very delicate issue of a reliable
{\it orbital-dependent} double-counting correction.

Another consequence of using delocalized Wannier functions is that
they lead to significant non-local interactions $V_{dd}$, which we found to be
(from cRPA) of order $0.23 U$ to
$0.32 U$. These interactions are completely neglected in the single-site
local DMFT approach, suggesting the need for a cluster extension in that case.
For these various reasons, we have reservations against using a $d$-only Hamiltonian with extended
Wannier functions for DMFT calculations on LaFeAsO,
as also previously emphasized Ref.~\onlinecite{vildosola1}.

\subsubsection{Consistency with previous calculations}

Nevertheless, in order to clarify apparent discrepancies between previously published
LDA+DMFT results\cite{haule1,anisimov2,haule2,shorikov1,anisimov3}, we performed
calculations within the $d$-hamiltonian, for several interaction parameters
reported in the literature.
For the values $U=4.0$\,eV and $J=0.7$\,eV used in Ref.~\onlinecite{haule1}, we
do confirm that the results then display very strong correlations
with quasiparticle renormalizations ranging from $Z=0.11$ ($xy$
orbital) to $Z=0.34$ ($x^2-y^2$ orbital).
One may note that
within the $d$-model there is a substantial orbital dependence of
$Z_m$, with a stronger renormalization predicted for
the $xy$, $yz$ and $zx$ orbitals. This is a clear consequence of
the Wannier functions being much more delocalized and anisotropic.

The scattering rate at this
inverse temperature of $\beta=40$\,eV$^{-1}$ is
quite sizable (${\rm Im}\Sigma(\omega^+=0)\approx -0.4\ldots-0.6$, depending
on the orbital), showing that the system is on the verge of a
coherence-incoherence crossover and a bad metal. The impurity spectral
function is plotted in Fig.~\ref{fig:laofeas_d}. It resembles very much the
one shown in Fig.~3 of
Ref.~\onlinecite{haule1}, showing clear signatures of lower and
upper Hubbard bands. There are, though, some discrepancies with the
total weight and the positions of the Hubbard bands, but given the
differences in the calculation (underlying electronic structure method,
temperature, interaction vertex which here is only density-density),
this agreement with Ref.~\onlinecite{haule1} is quite satisfactory.

Furthermore, using the parameters $U=0.8$\,eV and $J=0.5$\,eV from
Ref.~\onlinecite{anisimov2}, we find renormalizations in the
range $Z\approx 0.7 - 0.8$. This is somewhat smaller than reported
in Ref.\onlinecite{anisimov2}, although not in drastic disagreement.

Finally, we investigated the dependence on the Hund's rule coupling
of calculations performed with the $d$-only hamiltonian. Decreasing
$J$ to the much lower value
$J=0.2$\,eV but keeping $U=4$\,eV, we find the system to be
much less correlated ($Z$ between 0.63 and 0.73). We thus confirm,
for those calculations, the great sensitivity to the Hund's coupling
reported in Ref.~\onlinecite{haule2}.
We note however that, although reducing $J$ does make the system somewhat less
correlated in this case too, this sensitivity is much weaker when calculations are performed
with the full $dpp$ hamiltonian, as reported above.

\subsubsection{Origin of the sensitivity to the Hund's coupling: level crossings}

In order to understand the origin of the remarkable sensitivity of the
correlation
strength to the value of  $J$ observed with the $d$-Hamiltonian  we have studied
the evolution of the ground state of the Fe 3$d$ “atomic shell” as function of $J$.
We obtained the 3$d$ level positions corresponding to two different choices of the energy window:
the "small" one corresponding to the $d$-Hamiltonian and comprising 10 Fe 3$d$ bands and the “very large”
one comprising all As 4$p$, O 2$p$ and Fe 3$d$ bands as well as all unoccupied bands up to 13 eV
above $E_F$. The non-interacting level positions $\epsilon_{mm'}^{\alpha,\sigma}$ are then obtained as

\begin{equation}
\epsilon_{mm'}^{\alpha,\sigma}=\sum_{{\mathbf k},\nu\in \mathcal{W}} P_{m\nu}^{\alpha,\sigma}\epsilon_{\mathbf{k}\nu}^\sigma
P_{\nu m'}^{\alpha,\sigma *}-\tilde{\Sigma}_{mm'}^{\sigma,\rm{dc}},
\end{equation}
where the double counting term $\tilde{\Sigma}_{mm'}^{\sigma,\rm{dc}}$ is calculated in accordance
with Eq.~(\ref{eq:dc}) but with the "atomic" occupancy $N=6$ of the Fe 3$d$ shell. We used
the same values of $U=2.14$ and 2.69\,eV for the "small" and "very large" window choices, respectively,
while the value of $J$ was varied from 0.1 to 0.5\,eV. With $\epsilon_{mm'}^{\alpha,\sigma}$
corresponding to the $d$-Hamiltonian we observed a level crossing at $J \approx 0.2$ eV with the atomic ground state
changing from the one with spin moment $S=1$  to the one with $S=2$.
In the case of the "very large" window
the ground state always corresponds to $S=2$, and the splitting between the ground state and first excited level
is constant. It is obvious that a drastically different behavior of those two "atomic" models is related to
the corresponding level positions $\epsilon_{mm'}$, which are computed using different choices for the Wannier orbitals.
The observed change of the Fe 3$d$ atomic ground state, induced by increasing $J$, hints on
a possible strong dependence of correlation strength on the
Hund's rule coupling for LaFeAsO, which is indeed observed in our LDA+DMFT calculations with  the $d$-Hamiltonian.
However,
this sensitivity stems from a particular choice of delocalized and anisotropic Wannier functions
and is much less pronounced when the
energy window for the Wannier function construction is increased.

The bottom-line of this investigation is that all previously published
calculations seem to be technically correct. However, as discussed above, one
introduces several severe approximations when dealing with the
$d$-hamiltonian only, and the justification of these approximations
(restriction to local interactions, single-site DMFT, etc...)
is questionable. This is especially true in this compound, due to the
strong covalency between iron and arsenic states.

\section{Conclusion and prospects}

In the first part of this work, we present an implementation of
LDA+DMFT in the framework of the full-potential 
linearized augmented plane waves method.
We formulate the DMFT local impurity problem in the basis of Wannier orbitals, 
while the full lattice Green's function is written in the basis of Bloch eigenstates of the Kohn-Sham problem. In order to construct
the Wannier orbitals for a given correlated shell we choose a set of local orbitals, which are then expanded
onto the KS eigenstates lying within a certain energy window. In practice, 
we employ the radial solutions of the Schr\"odinger
equation for a given shell evaluated at the corresponding linearization energy as local orbitals. By orthonormalizing the obtained set of basis
functions we construct a set of true Wannier orbitals as well as projector operator matrices relating the Bloch and Wannier basis sets. We 
derive explicit formulas for the projected operator matrices in a
general FLAPW framework, which may include different types of
augmented plain waves, $lo$ and $LO$ orbitals. Our new implementation
is benchmarked using the test case of SrVO$_3$,  
for which we have obtained spectral and electronic properties in very good agreement with results of previous LDA+DMFT calculations.

In the second part of this paper we apply this LDA+DMFT technique to LaFeAsO in order to assess the degree of electronic
correlations in this compound and clarify the ongoing controversy about this issue in the literature. 
We solved the DMFT quantum impurity problem using a continuous-time quantum Monte Carlo approach.
The Wannier functions are constructed
using an energy window comprising Fe 3$d$, As 4$p$ and O 2$p$. The resulting Wannier orbitals are rather well localized and isotropic. 
We take the average 
values of $U=$2.69 eV and $J=$0.79 from constrained RPA calculations, where the Wannier functions and screening channels are consistent with our setting
of the LDA+DMFT scheme. We have checked the robustness of these results by increasing the size of the energy window, what resulted in a very similar 
physical picture.

Our LDA+DMFT results indicate that LaFeAsO is a moderately correlated metal with
an average value for the mass renormalization of the Fe 3$d$ bands  about 1.6. This value is in reasonable agreement 
with estimates from photoemission experiments. 	

We also consider a smaller energy window that includes Fe-$d$ states only. The  resulting Wannier functions in this case 
are quite extended, leading to anisotropic and non-local Coulomb interactions. 
We take different values for $U$ and $J$, including the ones used in previous theoretical LDA+DMFT approaches. We demonstrate 
that different physical pictures ranging from a strongly correlated compound on the verge of the metal insulator transition to a moderately
 to weakly correlated one can emerge depending, in particular, on the
 choice of the Hund's rule coupling $J$ as observed in Ref.~\onlinecite{haule2}. 
However, there are conceptual difficulties when constructing a local Hamiltonian from rather delocalized Wannier orbitals. The interactions 
are very anisotropic and orbital dependent, and non-local interactions could also become important. 
 
In summary, we demonstrate that the discrepancies
in the results of several recent theoretical works employing the LDA+DMFT approach stem from two main causes: i) the choice
of parameters of the local Coulomb interaction on the Fe 3$d$ shell and ii)
 the degree of localization of the Wannier orbitals chosen to represent the
Fe 3$d$ states, to which many-body terms are applied.
Regarding the first point, the calculated interaction parameters employed in the
present work  are significantly smaller than the values hypothesized
in Refs.~[\onlinecite{haule1,haule2}]. 
Regarding the second point, we provide strong evidence that the DMFT approximation is more accurate and more
straightforward to implement when well-localized orbitals are constructed
from a large energy window encompassing Fe-3$d$, As-4$p$ and O-2$p$.
This issue has fundamental implications for many-body calculations,
such as DMFT, in a realistic setting.

\appendix
\section{Benchmark: SrVO$_3$}\label{sect:benchmark}

For benchmarking purposes, we present in this appendix LDA+DMFT
results for an oxide that has become a classical test compound
for correlated electronic structure calculations, namely
the cubic perovskite \srvo3.
As a paramagnetic correlated metal with intermediate electron-electron
interactions, it is in a regime that is neither well described
by pure LDA calculations nor by approaches such as LDA+U that
are geared at ordered insulating materials.
From the experimental side, \srvo3 has been characterized
by different techniques (angle-resolved and angle-integrated
photoemission spectroscopy, optics, transport, thermodynamical
measurements etc)\cite{imada_mit_review,fujimori_pes_oxides,
ono91,maiti_2001,maiti_phd, inoue_casrvo3_1995_prl,sek04,
yos05,wad06-bis,sol06,egu06}.

LDA+DMFT calculations have been performed both for an effective low
energy model that comprises the three degenerate bands of
mainly \t2g character that are located around the Fermi
energy -- taking advantage of the cubic crystal field
that singles out this group of bands -- and for a bigger
energy window comprising also the oxygen $p$-states.
\cite{lie03,sek04,pav04,nek05,yos05,wad06-bis,sol06,nek06,
amadon_pw_08}.

In the low-energy effective \t2g model 
a quasi-particle renormalization of $Z \sim 0.6$,
compatible with experiments, is obtained for $U$ values around
4\,eV.
The remaining spectral weight is shifted toward lower and
upper Hubbard bands. The lower Hubbard band, located around
$-1.5$\,eV binding energy has indeed been observed in
photoemission; the high energy satellite of the \t2g
model is located around 2.5\,eV.
\footnote{Even though comparisons with x-ray absorption
spectroscopy have been attempted
it is not clear
that the low-energy description by a pure 
t2g model
is still valid at these energies. Note in particular
that in this energy region overlaps with the 
eg states,
split off by the cubic crystal field, could come into play.}

Concerning calculations taking into account also the ligand 
states, it should be noted that possible LDA errors on the
separation of $p$- and $d$-states are not corrected by DMFT,
since only the $d$-states are treated as correlated.

\begin{figure}[t]
  \centering
  \includegraphics[width=0.9\columnwidth]{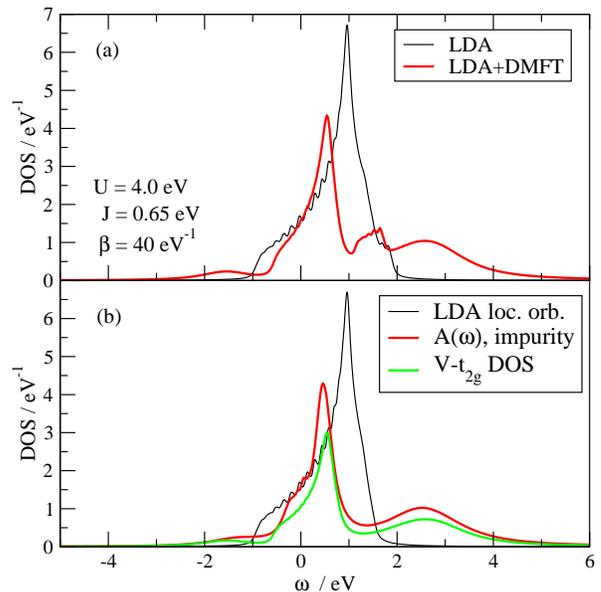}
  \caption{\label{fig:srvo3d}
    (Colore online) DOS for \srvo3, $d$-only model (small energy window). Top panel
    (a): Total DOS of LDA (black) and LDA+DMFT (red). Bottom panel
    (b): LDA local orbitals (black), impurity spectral function
    $A(\omega)$ (red), and vanadium \t2g partial DOS (green). Coulomb
    parameters for these calculations are given as inset.}
\end{figure}

\begin{figure}[t]
  \centering
  \includegraphics[width=0.9\columnwidth]{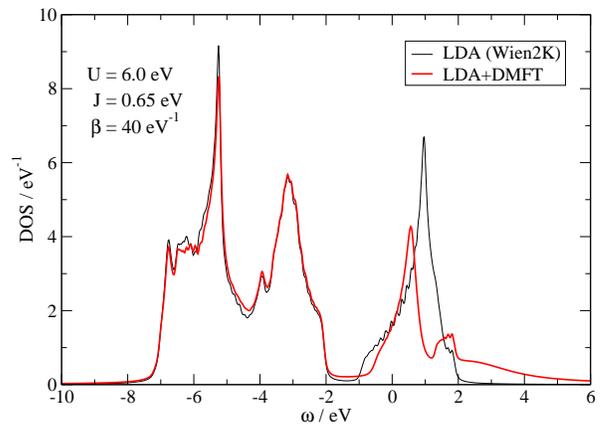}
  \caption{\label{fig:srvo3dp}
    (Color online) Total DOS for \srvo3, $dp$ Hamiltonian (vanadium \t2g and oxygen $p$).}
\end{figure}

In the present work, we use \srvo3 as a benchmark for our 
projector orbitals implementation of LDA+DMFT, with results 
very similar to previous theoretical studies.
We performed two kinds of calculations:
(i) We used as an energy window the range from $-1.35$\,eV to 1.90\,eV 
which comprises the \t2g bands, and -- at some k-points --
one or both of the \eg~bands.
This is closest in spirit to a \t2g model within a Wannier
function formalism, though not exactly the same due
to the inclusion of some \eg~contribution.
To recover a Wannier prescription one would in fact
have to choose a k-dependent window, such as to include
exactly three bands {\it at each k-point}, corresponding
to the three-fold degenerate manifold of dominantly
\t2g bands.
(ii) We used an energy window of $-8.10$\,eV to 1.90\,eV, spanning both, 
the bands used in (i) and the oxygen $p$ dominated
bands located between $-8$ and $-2$\,eV.

Please note that, in order to be consistent with existing literature,
we use a different parametrisation of the interaction matrix compared
to Sect.~\ref{sect:results}. Here we define $U$ to be the onsite
intraorbital Coulomb interaction, $U-2J$ to be the inter-orbital
interaction for electrons with opposite spin, and $U-3J$ the
inter-orbital interaction between electrons with equal spin.

The results for the first case are shown in Fig.~\ref{fig:srvo3d}.
The upper panel, Fig.~\ref{fig:srvo3d} (a), displays the total spectral
function of the thus defined model within LDA and LDA+DMFT.
Our results recover previously published results, with a quasiparticle
renormalisation of around $Z=0.60$ for a values $U=4.0$\,eV and $J=0.65$\,eV.
The contribution of the \eg~bands to the total DOS can easily be
identified from the LDA+DMFT spectra, where an additional hump between
the quasiparticle peak and the upper Hubbard band appears.

Fig.~\ref{fig:srvo3d} (b) shows the local orbitals used for the DMFT
calculations, together with the corresponding impurity spectral
function $A(\omega)$ and the vanadium \t2g partial DOS. The latter one
is obtained bu projecting the lattice Green's function to \t2g
character using the partial projectors to be introduced in
App.~\ref{sect:pc_proj}. The main difference to panel (a) is the
absence of the additional \eg~character.

Finally, Fig.~\ref{fig:srvo3dp} shows the LDA+DMFT spectral
function compared to the LDA density of states,
as calculated within the larger energy defined
in (ii) above. 
Since the Wannier function are more localised, as compared to case
(i), the value for the Coulomb interactions has to be adjusted
accordingly, and we chose a value of $U=6.0$\,eV.
As can be seen in the figure, ligand states are barely modified by the
correlations, and the results for the \t2g-derived bands are
very close to what is seen in the effective
low energy model. The quasiparticle renormalisation is $Z=0.57$, in
good agreement with the pure \t2g treatment discussed before.  

The results of these calculations correspond to what can be
expected on the basis of previously published work,
and thus validate our new implementation.

\label{sect:srvo3}

\section{Projectors for partial DOS}\label{sect:pc_proj}

In order to calculate the partial density of states for a given atomic site and particular orbital character 
 (correlated or not) we construct a different type of projectors, which we call $\hat{\bf{\Theta}}^{i,\sigma}$. 

The Wannier operators of Eq. \ref{eq:wannier-proj} project onto a given Wannier-like orbital. 
On the other hand, the new set $\hat{\bf{\Theta}}^{i,\sigma}$, as we will show, project onto a given orbital of certain 
character for which we do not apply any orthonormalization process as in the first. Unlike the Wannier projectors, the 
$\hat{\bf{\Theta}}^{i,\sigma}$'s can also project to other orbitals atoms apart from the correlated set. 

A given orbital character contributes in the eigenstates through the
solutions of the Schr\"{o}dinger equation inside the sphere $u_{l}^{\sigma}(r,E_{l1})Y_{m}^{l}\chi_{\sigma}$,
$\dot{u^{\sigma}}_{l}(r,E_{l1})Y_{m}^{l}\chi_{\sigma}$ and $u_{l}^{\sigma}(r,E_{l2})Y_{m}^{l}\chi_{\sigma}$
which do not form an orthonormalized basis set. It is more convenient
to construct these projectors if the wave function is rewritten in
an orthonormal basis set. 

In a general form, inside a given sphere we can express $\psi_{\mathbf{k}\nu}^{\sigma}(\mathbf{r})$
as 

\begin{equation}
\psi_{\mathbf{k}\nu}^{\sigma}(\mathbf{r})=\sum_{lm}A'_{lm}u_{l1}+\sum_{lm}B'_{lm}\dot{u}_{l}+\sum_{lm}C'_{lm}u_{l2},\label{eq:wave-func1}\end{equation}
where we simplify the notation by omitting the angular and spin parts
and defining $A'_{lm}$, $B'_{lm}$ and $C'_{lm}$ as combined coefficients
which are generally k-dependent and contain the sum over the plane
waves and local orbitals. We also define $u_{l1}\equiv$$u_{l}(r,E_{l1})$,
$\dot{u}_{l}\equiv\dot{u}_{l}(r,E_{l1})$ and $u_{l2}\equiv u_{l}(r,E_{l2})$.

We then rewrite $\psi_{\mathbf{k}\nu}^{\sigma}(\mathbf{r})$ as a
function of a set of orthogonal orbitals $\phi_{j}(r)$, \emph{j}=1,2,3
as follows

\begin{equation}
\psi_{\mathbf{k}\nu}^{\sigma}(\mathbf{r})=\sum_{lm}\sum_{j}\left(A'_{lm}c_{1j}^{lm}+B'_{lm}c_{2j}^{lm}+C'_{lm}c_{3j}^{lm}\right)\phi_{j}(r).\label{eq:wave-func2}\end{equation}

The coefficients $c_{ij}^{lm}$ are the matrix elements of the square
root of the corresponding overlap matrix

\begin{equation}
\mathbf{C}=\left(\begin{array}{ccc}
1 & 0 & \left\langle u_{l1}\right|\left.u_{l2}\right\rangle \\
0 & \left\langle \dot{u}_{l}\right|\left.\dot{u}_{l}\right\rangle  & \left\langle \dot{u}_{l}\right|\left.u_{l2}\right\rangle \\
\left\langle u_{l2}\right|\left.u_{l1}\right\rangle  & \left\langle u_{l2}\right|\left.\dot{u}_{l}\right\rangle  & \left\langle u_{l2}\right|\left.u_{l2}\right\rangle \end{array}\right)^{\frac{1}{2}}.\label{eq:overlap}\end{equation}

In this way, rewriting Eq. \ref{eq:wave-func2} as

\begin{equation}
\psi_{\mathbf{k}\nu}^{\sigma}(\mathbf{r})=\sum_{lm}\sum_{j}\tilde{c}_{j}^{lm}\phi_{j}(r).\label{eq:wave-func3}\end{equation}
the matrix elements of the projector to a given atom with \emph{lm}
character finally reads,

\begin{equation}
\Theta_{m\nu j}^{i,\sigma}(\mathbf{k})=\tilde{c}_{j}^{lm}.\label{eq:theta-proj-2}\end{equation}

The spectral function of a given atom \emph{i} with orbital character
\emph{m}, is obtained as

\[
A_{m}^{i,\sigma}(\mathbf{k},\omega)=-\frac{1}{\pi}Im\left[\sum_{\nu\nu',j}\Theta_{m\nu j}^{i,\sigma}(\mathbf{k})G_{\nu\nu'}^{\sigma}(\mathbf{k},\omega^{+})\Theta_{\nu'm'j}^{i,\sigma*}(\mathbf{k})\right]\]

\section{Influence of the rotational invariance of Hunds-rule coupling in multi-orbital systems}\label{sect:hundsrule}

\begin{figure}[t]
  \centering
  \includegraphics[width=0.8\columnwidth]{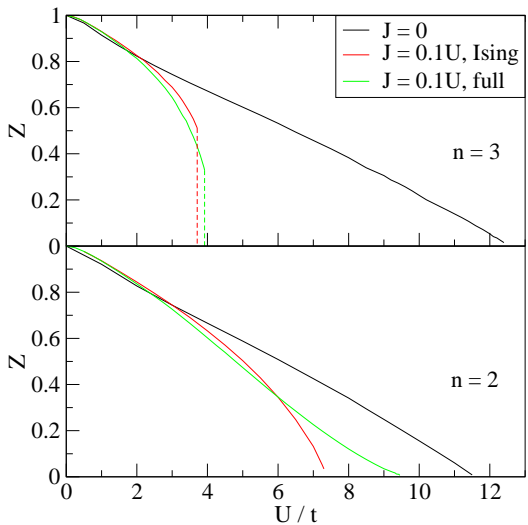}
  \caption{\label{figZ}
  Quasi-particle renormalisation $Z$ in a 3-orbital Hubbard
  model. Calculations have been done using a semicircular DOS with
  bandwidth $W=4t$. Top panel: $n=3$ (half filling). Bottom panel:
  $n=2$.} 
\end{figure}

In our DMFT calculations using CTQMC as impurity solver, we restricted
the Hund's rule interaction to Ising-type interactions only,
although there is no conceptual limitation of the algorithm to this
type of interactions. The
reason for doing this is of purely technical nature, since in this
case one can diagonalize the
local problem very efficiently, and furthermore, it enables us to use
the so-called segment-picture update scheme,\cite{werner2} which
increases the efficiency of the CTQMC method a lot. 

One may now ask how results change if the fully rotational-invariant
Hund's rule exchange is taken into account. For this purpose, we study
a multiband model Hamiltonian, assuming degenerate bands, no interband
hybridizations, and a
semicircular density of states. Applying the self-energy functional
theory (SFT),\cite{sft_potthoff} we can study the quasiparticle
renormalization $Z$ as function of interactions $U$ and $J$. In this
study, we choose the convention of setting the intraorbital Coulomb
repulsion to $U$ and the interorbital to $U^\prime=U-2J$, and give all
energies in units of the single-particle hopping amplitude $t$,
i.e., the band with of the DOS is $W=4t$. 

In addition to the density-density interactions, we consider also the
additional spin-flip and pair-hopping terms of the local Hamiltonian,
\begin{align}
        H_{\rm sf}&=-\frac{J}{2}\sum_{mm'}\left(c_{m\uparrow}^\dagger
        c_{m\downarrow} c_{m'\downarrow}^\dagger c_{m'\uparrow} + {\rm
        h.c.}\right) \\
        H_{\rm ph}&=-\frac{J}{2}\sum_{mm'}\left(c_{m\uparrow}^\dagger
        c_{m\downarrow}^\dagger c_{m'\uparrow} c_{m'\downarrow} + {\rm
        h.c.}\right). 
\end{align}
We do calculations at $T=0$, and choose the reference system for the
SFT framework to consist of one bath degree of freedom for each
correlated orbital. Hence, going up to $M=5$ orbitals, we have to
diagonalize a local problem consisting of at most 10 orbitals.

The upper panel of Fig.~\ref{figZ} shows $Z$ for a 3-orbital model at
half-filling, $n=3$, for $J=0.1U$. A tremendous reduction of the
critical $U_c$ of the metal-to-insulator transition (MIT) is observed,
already for Ising-like interactions. This is a well known fact
that for multi-orbital systems at or close to half-filling, the effect
of $J$ should be strongest.\cite{ono03,pruschke05,in.ko.05} 
The inclusion of spin-flip and pair-hopping terms gives raise to two
effects. (i) For moderate correlations, $Z\approx 0.6$, these terms
lead to a slight reduction of $Z$, but (ii) the critical $U$ for the
MIT is shifted upwards. This qualitatively holds also away from
half-filling, which can be seen in the lower panel of
Fig.~\ref{figZ}, where we plotted $Z$ for $n=2$. Although the
transition is not of first order any more, one can again identify two
regimes. For moderate correlations, $Z$ {\em decreases}, whereas close to
the transition the spin-flip and pair-hopping terms {\em
increase} the renormalization $Z$ and the critical $U_c$ is pushed to
higher values. This 
is consistent with a numerical renormalization group study for the 2-orbital Hubbard
model.\cite{pruschke05}

We also considered the case relevant for pnicitde materials,
i.e., $M=5$, $n=6$, and $Z$ around 0.5. This regime could be realize by
setting (i) $U=3.5t$ and $J=0.35t$, which shows a reduction of $Z$ from
0.52 to 0.47 due to spin-flip and pair-hopping, or (ii) $U=2t$,
$J=0.4t$, giving a reduction from 0.61 to 0.57. In conclusion, this
analysis shows that the picture of a moderately-correlated metal as
argued in Sect.~\ref{sect:results} holds also when a fully
rotational-invariant Hund's exchange is considered. For systems close
to a MIT, this is no longer true and the spin-flip and pair-hopping
terms become crucial.

\acknowledgments

We are grateful to Vladimir Anisimov, Ryotaro Arita, Gabriel Kotliar, Igor Mazin,
Frank Lechermann, Alexander Lichtenstein and, especially, Kristjan Haule for
useful discussions and correspondence. M.A. is grateful to F. Assaad
for enlightening discussions on the stochastic Maximum Entropy method. 
We acknowledge the support of the Agence Nationale de la Recherche (under project CORRELMAT), and
of GENCI and IDRIS (under project 091393) for supercomputer
time. M.A. acknowledges financial support from the Austrian Science
Fund (FWF), grant J2760-N16.

\end{document}